\documentclass[aps,pra,reprint,amsmath,amssymb,groupedaddress,showpacs]{revtex4-1}
\usepackage{graphicx}                  
\usepackage{float}
\usepackage{bm}                        
\usepackage[bookmarks=false]{hyperref} 
\usepackage[font={small}]{caption}     
\usepackage{enumitem}       
\usepackage{textcomp}       
\hypersetup{
    unicode=false,          
    pdftoolbar=true,        
    pdfmenubar=true,        
    pdffitwindow=true,      
    pdfstartview={},        
    pdftitle={Candidates for Universal Measures of Multipartite Entanglement}, 
    pdfauthor={Samuel R. Hedemann},                  
    pdfsubject={Multipartite Entanglement Measures}, 
    pdfcreator={},          
    pdfproducer={},         
    pdfkeywords={Entanglement,} {Multipartite Entanglement,} {Ent-Concurrence,} {Ent,} {GM Concurrence,} {k-Separability,} {Biseparability,} {TGX States,} {Absolute Ent-Concurrence}, 
    pdfnewwindow=true,      
    colorlinks=true,        
    linkcolor=black,        
    citecolor=black,        
    filecolor=black,        
    urlcolor=black          
}
\newcommand{\Sec}[1]{\hyperref[sec:#1]{Sec.{\kern 2pt}\ref*{sec:#1}}}
\newcommand{\Section}[1]{\hyperref[sec:#1]{Section~\ref*{sec:#1}}}
\newcommand{\Fig}[2][]{\hyperref[fig:#2]{Fig.{\kern 2pt}\ref*{fig:#2}#1}}
\newcommand{\Figs}[4]{\hyperref[fig:#1]{Figs.{\kern 2pt}\ref*{fig:#1}#2--\ref*{fig:#3}#4}}
\newcommand{\Figure}[2][]{\hyperref[fig:#2]{Figure~\ref*{fig:#2}#1}}
\newcommand{\Figures}[4]{\hyperref[fig:#1]{Figures~\ref*{fig:#1}#2--\ref*{fig:#3}#4}}
\newcommand{\App}[1]{\hyperref[sec:App.#1]{App.{\kern 2pt}\ref*{sec:App.#1}}}
\newcommand{\Appendix}[1]{\hyperref[sec:App.#1]{Appendix~\ref*{sec:App.#1}}}
\newcommand{\Eq}[1]{\hyperref[eq:#1]{(\ref*{eq:#1})}}
\newcommand{\Eqs}[2]{\hyperref[eq:#1]{(\ref*{eq:#1}--\ref*{eq:#2})}}
\newcommand{\Table}[2][]{\hyperref[tab:#2]{Table~\ref*{tab:#2}#1}}
\newcommand{\tr}[0]{\text{tr}}
\newcommand{\redx}[2]{{#1}{\kern -5.3pt}{{\textnormal{\raisebox{-1.2pt}{\scalebox{1.2}{\textasciicaron}}}}}{\kern -4.2pt}{~}^{(#2)}}
\newcommand{\mbar}[0]{\mathop {m}\limits^{{\kern -3.5pt}~_{\overline{{\kern 7pt}}}}}
\newcommand{\mbarsub}[0]{{\kern 0.0pt}\mathop {m}\limits^{{\kern -0.3pt}{\overline{{\kern 5.5pt}}}}{\kern 0.0pt}}
\newcommand{\nmaxnot}{n_{\,\overline{{\kern -1.8pt}\max^{~^{~^{~}}}\!\!\!\!\!\!\!\!\!\!}}\,}
\newcommand{\shiftmath}[2]{\textnormal{\raisebox{#1}[#1][#1]{$#2$}}}
\newcommand{\redxshift}[4]{{#1}{\kern -5.3pt}{{\textnormal{\raisebox{-1.2pt}{\scalebox{1.2}{\textasciicaron}}}}}{\kern -4.2pt}{~}^{{\kern #4}\shiftmath{#3}{(#2)}}}
\newcommand{\scalemath}[2]{\textnormal{\scalebox{#1}{$#2$}}}
\newcommand{\hsp}[1]{{\kern #1pt}}
\makeatletter
\newcommand{\mydotfill}{\leavevmode \cleaders \hb@xt@ .65em{\hss .\hss }\hfill \kern \z@}
\makeatother
\newcommand{\TOCone}[4]{\hyperlink{#1}{\textbf{#3}}&\hyperlink{#1}{\textbf{#4}}\mydotfill\!\! &\hspace{\stretch{1}}\hyperlink{#1}{\textbf{\pageref*{#2}}}}
\newcommand{\TOConeApp}[4]{\hyperlink{#1}{{\kern 7pt}\textbf{#3}}&\hyperlink{#1}{\textbf{#4}}\mydotfill\!\! &\hspace{\stretch{1}}\hyperlink{#1}{\textbf{\pageref*{#2}}}}
\newcommand{\TOCtwoApp}[4]{\hyperlink{#1}{{\kern 7pt}\textbf{#3}}&\hyperlink{#1}{\textbf{#4}}\mydotfill\!\! &\hspace{\stretch{1}}\raisebox{-10.8pt}{\hyperlink{#1}{\textbf{\pageref*{#2}}}}}
\newcommand{\TOCthrApp}[4]{\hyperlink{#1}{{\kern 7pt}\textbf{#3}}&\hyperlink{#1}{\textbf{#4}}\mydotfill\!\! &\hspace{\stretch{1}}\raisebox{-20.9pt}{\hyperlink{#1}{\textbf{\pageref*{#2}}}}}
\begin{document}
\title{Candidates for Universal Measures of Multipartite Entanglement}
\author{Samuel{\kern 3.0pt}R.{\kern 2.5pt}Hedemann}
\affiliation{P.O.{\kern 2.5pt}Box 72, Freeland, MD 21053, USA}
\date{\today}
\begin{abstract}
We propose and examine several candidates for universal multipartite entanglement measures.  The most promising candidate for applications needing entanglement in the full Hilbert space is the ent-concurrence, which detects all entanglement correlations while distinguishing between different types of distinctly multipartite entanglement, and simplifies to the concurrence for two-qubit mixed states. For applications where subsystems need internal entanglement, we develop the absolute ent-concurrence which detects the entanglement in the reduced states as well as the full state.
\end{abstract}
\maketitle
\section{\label{sec:I}Introduction}
\begin{figure}[H]
\centering
\vspace{-12pt}
\setlength{\unitlength}{0.01\linewidth}
\begin{picture}(100,0)
\put(1,24){\hypertarget{Sec:I}{}}
\end{picture}
\end{figure}
\vspace{-39pt}
Entanglement \cite{EPR1,Sch1}, in the simplest case of pure quantum states, is when a state such as $|\psi\rangle=a|1\rangle\otimes|1\rangle+b|1\rangle\otimes|2\rangle+c|2\rangle\otimes|1\rangle+d|2\rangle\otimes|2\rangle$, where $|a|^2+|b|^2+|c|^2+|d|^2 =1$, cannot be factored as a tensor product of pure states such as $|\psi\rangle\!=\!(w|1\rangle\!+\!{\kern 1pt}x|2\rangle)\otimes(y|1\rangle\!+\!{\kern 1pt}z|2\rangle)$, where $|w|^2+|x|^2 =1$ and $|y|^2+|z|^2 =1$, where we have expressed each qubit in a generic basis $\{|1\rangle,|2\rangle\}$ (our convention in this paper, and these kets do not imply Fock states \cite{Dira}).

\textit{Quantum mixed states} are \smash{$\rho\equiv\sum\nolimits_{j}p_{j}|\psi_j\rangle\langle\psi_{j}|$}, where \smash{$p_j\in[0,1]$}, \smash{$\sum\nolimits_{\shiftmath{1pt}{j}}p_{j}=1$}, and all \smash{$|\psi_{j}\rangle$} are pure.  For \textit{bipartite} systems, those composed of two subsystems (modes), mixed states \smash{$\varsigma\equiv\varsigma ^{(1,2)} $} are separable \textit{if and only iff} (iff)
\begin{equation}
\varsigma ^{(1,2)}  = \sum\nolimits_j {p_j \varsigma_j^{(1)}   \otimes\varsigma_j^{(2)}  } ,
\label{eq:1}
\end{equation}
where parenthetical superscripts are mode labels, and each \smash{$\varsigma_{j}^{\shiftmath{0pt}{{\kern 1.5pt}(m)}}$} is pure. Each mode-$m$ reduced state \smash{$\redx{\varsigma}{m}\!\equiv\!\tr_{\mbarsub}(\varsigma)$}, where \smash{$\mbar$} means ``not $m$'' (see \App{A}), admits a decomposition of the form \smash{$\redx{\varsigma}{m} \equiv \sum\nolimits_{\shiftmath{1pt}{j}} {p_j \redxshift{\varsigma}{m}{0pt}{0.5pt}_{{\kern -2pt}j} }$} \smash{$=\!\sum\nolimits_{\shiftmath{1pt}{j}} {p_j \varsigma_{j}^{\shiftmath{0pt}{{\kern 1.5pt}(m)}} }$} as proved in \App{B}, so if we only knew reductions{\kern 1.5pt} \smash{$\redx{\varsigma}{m}$}, we could search decompositions of each one to find the pair with matching sets \smash{$\{p_{j}\}$} such that \smash{$\varsigma ^{(1,2)} \!=\! \sum\nolimits_{\shiftmath{1pt}{j}} {p_j \redxshift{\varsigma}{1}{0pt}{0.5pt}_{{\kern -2pt}j}  \otimes \redxshift{\varsigma}{2}{0pt}{0.5pt}_{{\kern -2pt}j} }$}. Therefore knowledge of the reductions allows reconstruction of the \textit{parent state} $\varsigma$.

For $N$-partite ($N$-mode) systems, \textit{separability can occur in more than one way}.  For example, two different $3$-qubit pure states could have \textit{separable bipartitions} as $\rho=\redx{\rho}{1}\otimes\redx{\rho}{2,3}$ and $\varrho=\redx{\varrho}{1,2}\otimes\redx{\varrho}{3}$, so we call \textit{both} of them \textit{biseparable} or \textit{$2$-separable}, even though the mode groups that are separable for each state are different. 

These different mode-groupings are called \textit{partitions}, which are definitions of \textit{new modes} composed of (but not subdividing) the original modes $m_k$, as explained in \App{C}.  For example, a tripartite state like $\rho^{(1,2,3)}$ can have three unique bipartitions $\rho^{(1|2,3)},\rho^{(2|1,3)},\rho^{(3|1,2)}$ and one unique tripartition $\rho^{(1|2|3)}=\rho^{(1,2,3)}$, showing that in the absence of partitions, the commas \textit{are} the partitions.

\begin{figure}[H]
\centering
\vspace{1pt}
\includegraphics[width=0.87\linewidth]{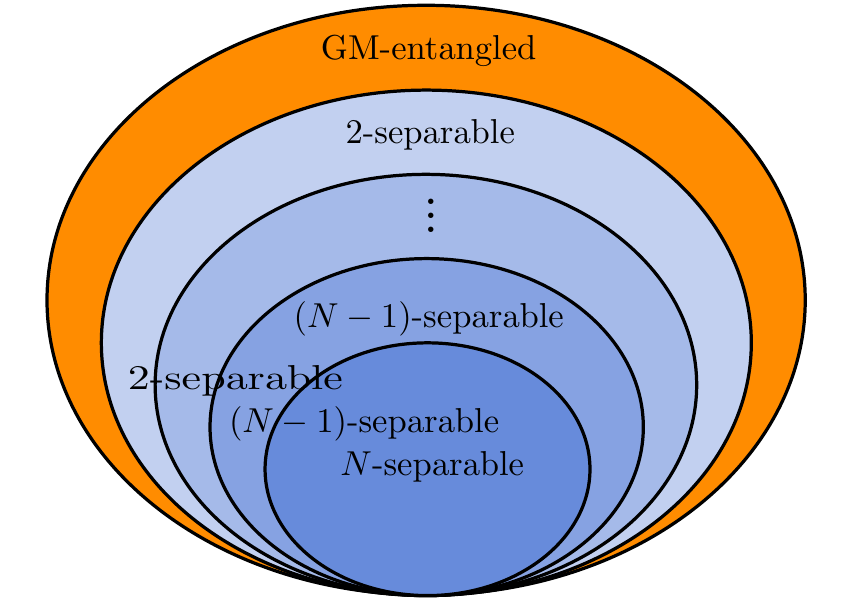}%
\vspace{-4pt}
\caption[]{(color online) Relationships of $k$-separabilities \cite{HSGS}. Each $k$-separability implies all lower-$k$-separabilities, and is necessary for all higher-$k$-separabilities. Thus, $N$-separable states are \textit{also} $(N-1)$-separable, all the way down to $2$-separable, but some $2$-separable states are not $3$-separable or higher. The ``1-separable'' states are ``genuinely multipartite (GM) entangled,'' also defined as all states that are not $2$-separable.  Thus, \textit{the GM-entangled region is strictly crescent-shaped here}, while \textit{the $k$-separable regions are each ellipse-shaped and coinciding with parts of all lower-$k$-separabilities}. (The shapes are arbitrary, merely representing relationships.)}
\label{fig:1}
\end{figure}

To handle the general $N$-partite phenomenon of separability of a given partitioning having the potential to occur in different ways, the notion of $k$-separability was developed \cite{HSGS,MCCS,HMGH,CoKW,Hor2,PlVe,EiBr,MeWa,SeSv,Miya,WoHo,YuSo,LOSU,HaJo}, as depicted in \Fig{1}.
\vspace{3pt}
By definition, $k$-separability of \textit{pure} states is when \textit{any} member of the set of all possible $k$-partitions has $k$-mode product form, for a fixed $k$. Thus, for a pure $\rho^{(1,2,3)}$, if only $\rho^{(2|1,3)}$ is $2$-separable, but not $\rho^{(1|2,3)}$ or $\rho^{(3|1,2)}$, then that is sufficient for $\rho^{(1,2,3)}$ to be $2$-separable.
\vspace{2pt}

\textit{Mixed} states are $k$-separable if a decomposition exists for which all pure decomposition states are \textit{at least} $k$-separable, with one being exactly $k$-separable \cite{HSGS} (since higher-than-$k$-separabilities are also $k$-separable, we can just say that all decomposition states need to be $k$-separable). For example, the $3$-qudit state
\begin{equation}
\begin{array}{*{20}l}
   {\rho} &\!\! {= \sum\limits_j {{\kern -1pt}p_j \rho _j^{(1)} \! \otimes{\kern -1pt} \rho _j^{(2,3)} } \! +\! \sum\limits_k {{\kern -1pt}q_k \rho _k^{(2)} \! \otimes{\kern -1pt} \rho _k^{(1,3)} } \! +\! \sum\limits_l {{\kern -1pt}r_l \rho _l^{(3)} \! \otimes{\kern -1pt} \rho _l^{(1,2)} }\!,}  \\
\end{array}
\label{eq:2}
\end{equation}
where \smash{$p_j ,q_k ,r_l  \in [0,1]$}, \smash{$(\sum\nolimits_j {p_j } )\! +\! (\sum\nolimits_k {q_k } ) \! +\! (\sum\nolimits_l {r_l } ) = 1$}, with pure entangled bipartite states \smash{$\rho _{{\kern -0.3pt}j}^{\shiftmath{-1pt}{{\kern 2pt}(2,3)}} ,\rho _{{\kern -0.3pt}k}^{\shiftmath{-1pt}{{\kern 2pt}(1,3)}} ,\rho _{{\kern -0.3pt}l}^{\shiftmath{-1pt}{{\kern 2pt}(1,2)}} $} and pure \smash{$\rho _{{\kern -0.5pt}\shiftmath{0.5pt}{j}}^{{\kern 1.5pt}\shiftmath{0.5pt}{(1)}} ,\rho _k^{{\kern 1.5pt}\shiftmath{0.5pt}{(2)}} ,\rho _l^{{\kern 1.5pt}\shiftmath{0.5pt}{(3)}} $}, is \textit{$2$-separable} (\textit{biseparable}), even though each group of pure decomposition states is separable over \textit{different} bipartitions \cite{HSGS}.

\begin{figure}[H]
\centering
\includegraphics[width=0.87\linewidth]{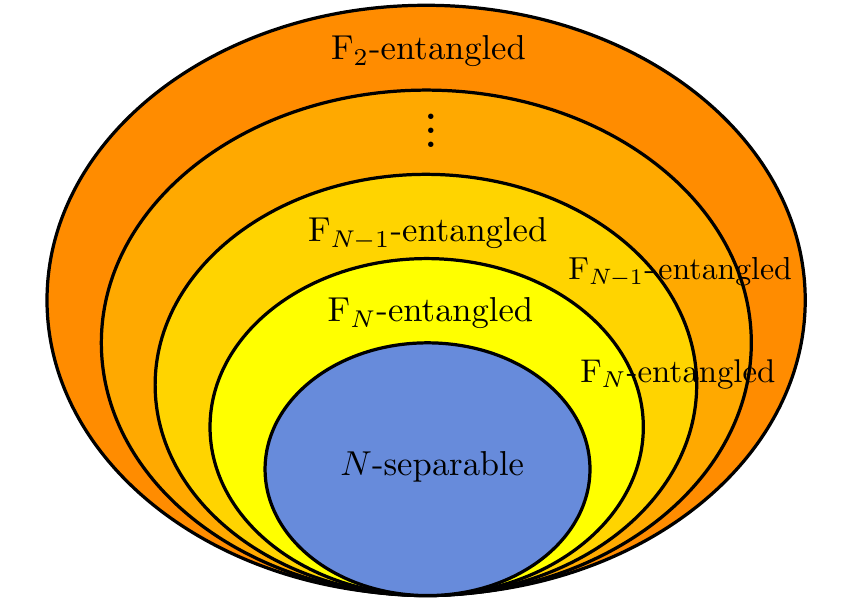}%
\vspace{-4pt}
\caption[]{\hsp{-2.5}(color\hsp{-0.6} online)\hsp{-1} Relationships\hsp{-0.1} of\hsp{-0.1} $\text{F}_k$-entanglements.\hsp{-1.6} Here,\hsp{-0.9} a\hsp{-0.9} given\hsp{-0.9} $\text{F}_k$-entanglement\hsp{-0.9} implies\hsp{-0.9} all\hsp{-0.9} higher-$k$\hsp{-0.9} $\text{F}_k$-entanglements, while being itself necessary for all lower-$k$ $\text{F}_k$-entanglements. Thus, \textit{each} $\text{F}_k$-\textit{entanglement region is strictly crescent-shaped and\hsp{-0.9} coinciding\hsp{-0.9} with\hsp{-0.9} parts\hsp{-0.9} of\hsp{-0.9} all\hsp{-0.9} higher-$k$}\hsp{-0.9} $\text{F}_k$-\textit{entanglements},\hsp{-0.9} with $\text{F}_N$-entanglement being the thickest crescent, and all lower-$k$ $\text{F}_k$-entanglements having progressively thinner crescents. The entire region that is not $N$-separable is $\text{F}_N$-entangled.}
\label{fig:2}
\end{figure}

Here,\hsp{-0.7} we\hsp{-0.7} define\hsp{-0.7} the\hsp{-0.7} ``absence\hsp{-0.7} of\hsp{-0.7} $k$-separability''\hsp{-0.7} (meaning no $k$-partitions of a pure state have $k$-partite product form) as \textit{full $k$-partite entanglement} ($\text{F}_k$-\textit{entanglement}) for $k\hsp{-2}>\hsp{-2}1$ as shown in \Fig{2}, found by \textit{combining} entanglement values (by some measure) over all $k$-partitions, and named in analogy to ``full $N$-partite entanglement'' being the absence of full $N$-partite separability.  However, the absence of $2$-separability is \textit{also} ``$2$-entanglement,'' known as ``genuinely multipartite (GM) entanglement.'' We use the term \textit{GM-$k$-entanglement} ($\text{GM}_k$-\textit{entanglement}) in lieu of the traditional term ``$k$-entanglement'' as the minimum entanglement over all $k$-partitions.
\vspace{-2pt}
\vspace{-7pt}
On the other hand, $N$-partite states are entangled iff they are $\text{F}_N$-entangled, as proved in \App{D}.  This means that the complete absence of entanglement correlations can only occur in $N$-partite states that are $N$-separable, meaning states with an \textit{optimal decomposition},
\vspace{-3pt}
\begin{equation}
\varsigma ^{(1, \ldots ,N)}  \hsp{-1}=\hsp{-2} \sum\nolimits_j {\hsp{-1}p_j \hsp{-2} \mathop  \otimes \limits_{m = 1}^N \hsp{-2}\varsigma_j^{(m)} } \hsp{-1} = \hsp{-2}\sum\nolimits_j {\hsp{-1}p_j \varsigma_j^{(1)} \hsp{-2}  \otimes  \cdots  \otimes \varsigma_j^{(N)}  } \hsp{-2},
\label{eq:3}
\end{equation}
where\hsp{-1.4} the\hsp{-0.8} \smash{$\varsigma_{j}^{(m)}$}\hsp{-2.1} are\hsp{-1.4} pure.\hsp{-1.7} $N$-partite\hsp{-1.4} states\hsp{-1.4} that\hsp{-1.4} \textit{cannot}\hsp{-1.4} be\hsp{-1.4} expanded\hsp{-1} as\hsp{-1} \Eq{3}\hsp{-1} are\hsp{-1} \textit{full\hsp{-1} $N$-partite\hsp{-1} entangled}\hsp{-1} ($\text{F}_N$-\textit{entangled}).\hsp{-1} We will often say \textit{entanglement correlation} instead of just \textit{entanglement} to remind that there are \textit{other} kinds of nonlocal correlation not involving entanglement.

However,\hsp{-1} $\text{F}_N$-entanglement\hsp{-1} cannot\hsp{-1} distinguish\hsp{-1} between\hsp{-1} types\hsp{-1} of\hsp{-1} multipartite\hsp{-1} entanglement.\hsp{-1} For\hsp{-1} example,\hsp{-1} given
\vspace{-3pt}
\begin{equation}
\rho _{|\Phi_{\text{GHZ}} \rangle } \;\;\;\text{and}\;\;\; \rho _{|\Phi_{\text{BP}} \rangle }  \equiv\rho _{|\Phi ^ +  \rangle }  \otimes \rho _{|\Phi ^ +  \rangle },
\label{eq:4}
\end{equation}
where \smash{$|\Phi_{\shiftmath{0.5pt}{\text{GHZ}}} \rangle \! \equiv\! \shiftmath{1pt}{\textstyle{1 \over {\sqrt 2 }}}(|1,\!1,\!1,\!1\rangle \! +\! |2,\!2,\!2,\!2\rangle )$} is a $4$-qubit GHZ state \cite{GHZ,GHSZ,Merm} where \smash{$\rho _{\shiftmath{0.5pt}{|A\rangle} }  \!\equiv\! |A\rangle \langle A|$}, and \smash{$|\Phi ^ +  \rangle \! \equiv\! \shiftmath{1pt}{\textstyle{1 \over {\sqrt 2 }}}(|1,1\rangle  +$} \smash{$ |2,2\rangle )$} is a $2$-qubit Bell state so that \smash{$\rho _{|\Phi_{\text{BP}} \rangle }$} is a ``Bell-product state,'' since both \smash{$\rho _{|\Phi_{\text{GHZ}} \rangle }$} and \smash{$\rho _{|\Phi_{\text{BP}} \rangle }$} have maximal mixing in all single-mode reductions, an $\text{F}_N$-entanglement measure would report both states as being equally entangled, despite \smash{$\rho _{|\Phi_{\text{GHZ}} \rangle }$} being $\text{F}_2$-entangled and $\text{GM}_2$-entangled while \smash{$\rho _{|\Phi_{\text{BP}} \rangle }$} is $2$-separable.

Yet, $\text{GM}_2$-entanglement\rule{0pt}{10pt} alone cannot detect the strong entanglement correlations within the Bell states of \smash{$\rho _{|\Phi_{\text{BP}} \rangle }$} in \Eq{4}.  Thus, while $\text{F}_N$-entanglement can detect the presence of all entanglement correlations but cannot distinguish $k$-separabilities, lone sub-$N$ $\text{GM}_k$-entanglement measures are not sufficient to detect the presence of all entanglement correlations, but \textit{can} verify $k$-separability.

Therefore\hsp{-0.5} our\hsp{-0.5} main\hsp{-0.5} goal\hsp{-0.5} here\hsp{-0.5} is\hsp{-0.5} to\hsp{-0.5} define\hsp{-0.5} a\hsp{-0.5} few\hsp{-0.5} candidate\hsp{-0.5} universal\hsp{-0.5} entanglement\hsp{-0.5} measures\hsp{-0.5} that\hsp{-0.5} distinguish\hsp{-0.5} between\hsp{-0.5} types\hsp{-0.5} of\hsp{-0.5} multipartite\hsp{-0.5} entanglement\hsp{-0.5} without\hsp{-0.5} discarding\hsp{-0.5} information\hsp{-0.5} about\hsp{-0.5} entanglement\hsp{-0.5} correlations,\hsp{-0.5} which\hsp{-0.5} individual\hsp{-0.4} $\text{GM}_k$-entanglement\hsp{-0.4} measures\hsp{-0.4} cannot\hsp{-0.4} do\hsp{-0.4} alone.

The building-block of our candidate measures is the $\text{F}_N$-entanglement measure \textit{the ent} \cite{HedE}, given by
\begin{equation}
\Upsilon (\rho) \equiv{\kern -1pt}\Upsilon(\rho,\mathbf{n}) {\kern -1pt}\equiv{\kern -1pt} \frac{1}{M(L_{*})}{\kern -1pt}\left({\kern -1pt} {1 - \frac{1}{N}\sum\limits_{m = 1}^N \!{\frac{{n_m P(\redx{\rho}{m}) - 1}}{{n_m  - 1}}} }{\kern -1pt} \right)\!,
\label{eq:5}
\end{equation}
for pure states $\rho$ of an $N$-mode $n$-level system where mode $m$ has $n_{m}$ levels and $\mathbf{n}\equiv(n_{1},\ldots, n_{N})$, so that $n=n_{1}\cdots n_{N}=\text{det}(\text{diag}(\mathbf{n}))$, $N=\text{dim}(\mathbf{n})$, $P(\sigma ) \equiv \tr(\sigma ^2 )$ is the purity of $\sigma$, and \smash{$\redx{\rho}{m}$} is the $n_m$-level single-mode reduction of $\rho$ for mode $m$ (see \App{A}). The normalization factor $M(L_{*})\equiv M(L_{*},\mathbf{n})$ is given in \App{E}.  Basically, the ent measures how simultaneously mixed the \smash{$\redx{\rho}{m}$} are.

We will also use the \textit{partitional ent} \smash{$\Upsilon^{(\mathbf{m}^{\shiftmath{-1pt}{(\mathbf{T})}})}(\rho)\equiv$} \smash{$\Upsilon(\redx{\rho}{\mathbf{m}},\mathbf{n}^{(\mathbf{m}^{\shiftmath{-1pt}{(\mathbf{T})}})})$}, allowing us to repartition $\rho$'s reduction \smash{$\redx{\rho}{\mathbf{m}}$} (including the nonreduction \smash{$\redx{\rho}{\mathbf{N}}\equiv\redx{\rho}{1,\ldots,N}=\rho$}) into new mode groups \smash{$\mathbf{m}^{(\mathbf{T})}\equiv(\mathbf{m}^{(1)}|\ldots|\mathbf{m}^{(T)})$} of levels \smash{$\mathbf{n}^{(\mathbf{m}^{\shiftmath{-1pt}{(\mathbf{T})}})}\equiv(n_{\mathbf{m}^{(1)}},\ldots,n_{\mathbf{m}^{(T)}})$} (see \App{C}) to measure the $\text{F}_T$-entanglement of any $T$ mode groups (see \cite{HedE} for details). Finally,{\kern 0.8pt} when{\kern 0.8pt} $\rho${\kern 0.8pt} or{\kern 0.8pt} \smash{$\redx{\rho}{\mathbf{m}}$} are \textit{mixed}, we use convex-roof{\kern -0.5pt} extensions{\kern -0.5pt} as{\kern -0.5pt} \smash{$\hat{\Upsilon}$}{\kern -0.5pt} and{\kern -0.5pt} \smash{$\hat{\Upsilon}^{(\mathbf{m}^{\shiftmath{-1pt}{(\mathbf{T})}})}$} (see \hyperlink{HedE}{\cite*[App.{\kern 2.5pt}J]{HedE}}), which are minimum average ents over all decompositions. The main sections of this paper are:
\vspace{-4pt}
\begin{table}[H]
{\begin{tabular}[b]{l@{\kern 2.5pt}p{0.83\linewidth}@{}p{0.05\linewidth}}
\TOCone{Sec:I}{sec:I}{I.}{Introduction}\\
\TOCone{Sec:II}{sec:II}{II.}{Candidate{\kern -1.6pt}\scalebox{0.95}[1]{ Pure-State{\kern -1.6pt} Entanglement{\kern -1.6pt} Measures}}\\
\TOCone{Sec:III}{sec:III}{III.}{Tests of Candidate Pure-State Measures}\\
\TOCone{Sec:IV}{sec:IV}{IV.}{Candidate Mixed-State Measures}\\
\TOCone{Sec:V}{sec:V}{V.}{Tests of Candidate Mixed-State Measures}\\
\TOCone{Sec:VI}{sec:VI}{VI.}{Ent-Concurrence}\\
\TOCone{Sec:VII}{sec:VII}{VII.}{Absolute Ent-Concurrence}\\
\TOCone{Sec:VIII}{sec:VIII}{VIII.}{Conclusions}\\
\TOCone{Sec:IX}{sec:App.A}{App.}{Appendices}\\
\TOConeApp{Sec:App.A}{sec:App.A}{A.}{Brief Review of Reduced States}\\
\TOCtwoApp{Sec:App.B}{sec:App.B}{B.}{$N$-Separability of $N$-Partite States Implies Reconstructability by Smallest Reductions}\\
\TOConeApp{Sec:App.C}{sec:App.C}{C.}{Definition of Partitions}\\
\TOCtwoApp{Sec:App.D}{sec:App.D}{D.}{Proof that $N$-Partite States are Entangled If and Only If they are $\text{F}_N$-Entangled}\\
\TOConeApp{Sec:App.E}{sec:App.E}{E.}{Normalization Factor of the Ent}\\
\TOConeApp{Sec:App.F}{sec:App.F}{F.}{True-Generalized X (TGX) States}\\
\TOCtwoApp{Sec:App.G}{sec:App.G}{G.}{Full Set of 4-Qubit Maximally $\text{F}_N$-Entangled TGX States Involving $|1\rangle$}\\
\TOCtwoApp{Sec:App.H}{sec:App.H}{H.}{Quantum{\kern -1.1pt} Mixed{\kern -1.1pt} States{\kern -1.1pt} Cannot{\kern -1.1pt} Be{\kern -1.1pt} Treated{\kern -1.1pt} as Time-Averages of Varying Pure States}\\
\end{tabular}}{\kern -10pt}
\end{table}{\kern -10pt}
{\noindent}and wherever possible, details are put in the \hyperlink{Sec:App.A}{Appendices}.
\section{\label{sec:II}Candidate Pure-State Entanglement Measures}
\begin{figure}[H]
\centering
\vspace{-12pt}
\setlength{\unitlength}{0.01\linewidth}
\begin{picture}(100,0)
\put(1,27){\hypertarget{Sec:II}{}}
\end{picture}
\end{figure}
\vspace{-39pt}
Here we introduce the candidate entanglement measures under consideration. Each will use the ent from \Eq{5} as well as its various different forms due to partitioning.  See \cite{HedE} for full explanations. We start with \textit{pure}-state measures, and discuss mixed input after initial tests.
\subsection{\label{sec:II.A}Full Genuinely Multipartite (FGM) Ent}
The FGM ent for pure $\rho$ is
\begin{equation}
\Upsilon _{\text{FGM}} (\rho ) \equiv \frac{1}{M_{\text{FGM}}}\sum\limits_{k = 2}^N {\Upsilon_{\text{GM}_k } (\rho )},
\label{eq:6}
\end{equation}
where \smash{$M_{\text{FGM}}$} is a normalization factor, and
\begin{equation}
\Upsilon _{\text{GM}_{k}} (\rho ) \equiv \min (\{ \Upsilon ^{(\mathbf{N}_h^{(\mathbf{k})} )}(\rho ) \}),
\label{eq:7}
\end{equation}
which is the $\text{GM}_k$ ent, where \smash{$\{ \Upsilon ^{(\mathbf{N}_h^{(\mathbf{k})} )}(\rho ) \}$} is the set of all $N$-mode $k$-partitional ents, each labeled by $h$. Note that \cite{HedE} defined the \textit{GM ent} as $\Upsilon_{\text{GM}}(\rho)\equiv\Upsilon_{\text{GM}_2}(\rho)$, which is the ``ent-version'' of GM concurrence $C_{\text{GM}}(\rho)$ \cite{MCCS}.

Since{\kern -1pt} \smash{$\Upsilon _{\text{FGM}}$}{\kern -1pt} sums{\kern -1pt} all{\kern -1pt} $\text{GM}_k${\kern -1pt} ents,{\kern -1pt} it{\kern -1pt} is{\kern -1pt} a{\kern -1pt} measure of \textit{simultaneous} $\text{GM}_k$-\textit{entanglements}; that is, it rates states for which the \textit{combination} of all their $\text{GM}_k$-entanglements is maximal as being ``maximally FGM-entangled.''
\subsection{\label{sec:II.B}Full Simultaneously Multipartite (FSM) Ent}
The FSM ent for pure $\rho$ is
\begin{equation}
\Upsilon _{\text{FSM}} (\rho ) \equiv \frac{1}{M_{\text{FSM}}}\sum\limits_{k = 2}^N {\Upsilon _{\text{SM}_{k}} (\rho ) },
\label{eq:8}
\end{equation}
where \smash{$M_{\text{FSM}}$} is a normalization factor, and we define the \textit{simultaneously multipartite $k$-ent} (\smash{$\text{SM}_{k}$} ent) as
\begin{equation}
\Upsilon _{\text{SM}_{k}} (\rho ) \equiv \frac{1}{M_{\text{SM}_{k}}}\sum\nolimits_{h = 1}^{\{_{\,k}^{N}\}} \Upsilon^{(\mathbf{N}_h^{(\mathbf{k})} )}(\rho ),
\label{eq:9}
\end{equation}
where\hsp{-1} \smash{$\{_{{\kern 1pt}k}^{N}\}\equiv\frac{1}{k!}\sum\nolimits_{{\kern 0.5pt}\shiftmath{1.4pt}{j\! =\! 0}}^k {\!( - 1{\kern -0.5pt})^{k - j} \binom{k}{j}j^N }$}\hsp{-1} are\hsp{-0.5} Stirling\hsp{-0.5} numbers of\hsp{-0.7} the\hsp{-0.7} second\hsp{-0.7} kind,\hsp{-1.7} \smash{$\binom{\shiftmath{-1pt}{k}}{\shiftmath{1pt}{j}}\hsp{-0.5}\equiv\hsp{-0.5}\frac{k!}{j!(k-j)!}$},\hsp{-0.7} \smash{$\{ \Upsilon^{\shiftmath{-0.8pt}{(\mathbf{N}_{{\kern -0.5pt}\shiftmath{-0.5pt}{h}}^{{\kern 0.5pt}\shiftmath{-0.3pt}{(\mathbf{k})}} )}}(\rho ) \}$}\hsp{-0.7} is\hsp{-0.7} the\hsp{-0.7} set\hsp{-0.7} of all $N$-mode $k$-partitional ents, and \smash{$M_{\text{SM}_{k}}$} is a normalization factor. \smash{$\Upsilon _{\text{SM}_{k}} (\rho )$} detects the presence of \textit{any} entanglement correlation among \textit{all} possible $k$-partitions of $\rho$.  Thus, it cannot ignore entanglement within particular $k$-partitions just because a different $k$-partition is separable as \smash{$\Upsilon _{\text{GM}_{k}} (\rho )$} can.

The FSM ent \smash{$\Upsilon _{\text{FSM}}$} is a measure of \textit{simultaneous} $\text{SM}_{k}$ ents; which means it is a measure of the combination of \textit{all} $N$-mode partitional ents, so it is a sum of all 2-partitional ents of $\rho$, all 3-partitional ents of $\rho$, all the way up to the $N$-partitional ent (the ent itself).  Thus, if there is \textit{any} entanglement correlation between \textit{any} mode groups of $\rho$, the FSM will detect it, and states that maximize it are ``maximally FSM-entangled.''
\subsection{\label{sec:II.C}Full Distinguishably Multipartite (FDM) Ent: The Ent-Concurrence}
The FDM ent (or the \textit{ent-concurrence}) for pure $\rho$ is
\begin{equation}
\Upsilon _{\text{FDM}} (\rho ) \equiv \frac{1}{M_{\text{FDM}}}\sum\limits_{k = 2}^N {\Upsilon _{\text{DM}_{k}} (\rho ) },
\label{eq:10}
\end{equation}
where \smash{$M_{\text{FDM}}$} is a normalization factor, and we define the \textit{distinguishably multipartite $k$-ent} (\smash{$\text{DM}_k$} ent) as
\begin{equation}
\Upsilon _{\text{DM}_{k}} (\rho ) \equiv \frac{1}{M_{\text{DM}_{k}}}\sum\nolimits_{h = 1}^{\{_{\,k}^{N}\}} \sqrt{\Upsilon^{(\mathbf{N}_h^{(\mathbf{k})} )}(\rho )},
\label{eq:11}
\end{equation}
where \smash{$M_{\text{DM}_{k}}$} is a normalization factor, \smash{$\{_{{\kern 1pt}k}^{N}\}$} are Stirling numbers of the second kind as in \Eq{9}, and \smash{$\{ \Upsilon^{\shiftmath{-0.8pt}{(\mathbf{N}_{{\kern -0.4pt}\shiftmath{-0.5pt}{h}}^{{\kern 0.5pt}\shiftmath{-0.3pt}{(\mathbf{k})}} )}}(\rho ) \}$} is the set of all $N$-mode $k$-partitional ents.

The FDM ent \smash{$\Upsilon _{\text{FDM}}$} is a measure of simultaneous \smash{$\text{DM}_k$} ents \smash{$\Upsilon _{\text{DM}_{k}}$}, and the \smash{$\Upsilon _{\text{DM}_{k}}$} measures not only the combination of all possible $N$-mode $k$-partitional ents, but also \textit{how equally distributed} they are, rating states for which all $N$-mode $k$-partitional ents have the highest combination \text{and} are the most equal and numerous as having the highest \smash{$\text{DM}_k$} ent. 

This{\kern -0.5pt} is{\kern -0.5pt} based{\kern -0.5pt} on{\kern -0.5pt} \textit{pseudonorm}{\kern -0.5pt} $|\mathbf{x}|_{1/2}\equiv\sum\nolimits_{k}\!\sqrt{x_{k}};\,x_{k}\geq 0$, which obeys $|\mathbf{x}|_{1/2}\!=\!0 \Rightarrow \mathbf{x}\!=\!\mathbf{0}$ and the triangle inequality, and although $|a\mathbf{x}|_{1/2}\neq|a|\!\cdot\!|\mathbf{x}|_{1/2}$, that does not matter here, since our ``vectors'' are really just lists of scalars. The main reason we use this pseudonorm is because it rates an equally distributed $1$-norm-normalized vector such as $\mathbf{x}=(0.25,0.25,0.25,0.25)$ as having the \textit{highest} ``$\frac{1}{2}$-norm'' value out of all vectors of the same $1$-norm, such as $\mathbf{y}=(0,0,0.5,0.5)$ or $\mathbf{z}=(0,1,0,0)$.

For two qubits ($N=2$), the FDM ent is the \textit{concurrence} $C$ \cite{HiWo,Woot}, since $\Upsilon=C^2$ as proven in \cite{HedE}, so that
\begin{equation}
\Upsilon _{\text{FDM}} (\rho )=\Upsilon _{\text{DM}_{N}} (\rho )=\sqrt{\Upsilon(\rho)}=C(\rho),
\label{eq:12}
\end{equation}
(extendible to mixed states, as shown later). Therefore, if we let the \textit{$N$-mode $k$-partitional ent-concurrence} be 
\begin{equation}
C_{\Upsilon}^{(\mathbf{N}_h^{(\mathbf{k})} )} (\rho )\equiv\sqrt{\Upsilon^{(\mathbf{N}_h^{(\mathbf{k})} )}(\rho )},
\label{eq:13}
\end{equation}
where \smash{$\{ \Upsilon^{(\mathbf{N}_h^{(\mathbf{k})} )}(\rho ) \}$} is the set of all $N$-mode $k$-partitional ents, then the \smash{$\text{DM}_k$} ent \smash{$\Upsilon _{\text{DM}_{k}}$} is simply a \textit{$1$-norm} of all \smash{$\{ C_{{\kern -1pt}\shiftmath{-0.9pt}{\scalemath{0.85}{\Upsilon}}}^{{\kern 0.0pt}\shiftmath{-0.4pt}{(\mathbf{N}_{{\kern -0.5pt}h}^{{\kern 0.5pt}\shiftmath{-0.3pt}{(\mathbf{k})}} )}}(\rho ) \}$} for a given $k$.  Therefore, the \smash{$\Upsilon _{\text{FDM}}$} in \Eq{10} can also be called the \textit{ent-concurrence} \smash{$C_{\Upsilon}$} for pure states.
\section{\label{sec:III}Tests of Candidate Pure-State Measures}
\begin{figure}[H]
\centering
\vspace{-12pt}
\setlength{\unitlength}{0.01\linewidth}
\begin{picture}(100,0)
\put(1,27){\hypertarget{Sec:III}{}}
\end{picture}
\end{figure}
\vspace{-39pt}
For simple tests, we use $4$-qubit ($2\times 2\times 2\times 2$) states,
\begin{equation}
\begin{array}{*{20}l}
   {|\Phi _{\text{GHZ}} \rangle } &\!\! { \equiv {\textstyle{1 \over {\sqrt 2 }}}(|1,\!1,\!1,\!1\rangle  \!+\! |2,\!2,\!2,\!2\rangle )}  \\
   {|\Phi _{\text{BP}} \rangle } &\!\! { \equiv\rule{0pt}{8.5pt} {\textstyle{1 \over {\sqrt 4 }}}(|1,\!1,\!1,\!1\rangle  \!+\! |1,\!1,\!2,\!2\rangle  \!+\! |2,\!2,\!1,\!1\rangle \! +\! |2,\!2,\!2,\!2\rangle )}  \\
   {|\Phi _{\text{F}} \rangle } &\!\! { \equiv\rule{0pt}{8.5pt} {\textstyle{1 \over {\sqrt 4 }}}(|1,\!1,\!1,\!1\rangle  \!+\! |1,\!1,\!2,\!2\rangle \! +\! |2,\!2,\!1,\!2\rangle  \!+\! |2,\!2,\!2,\!1\rangle )}  \\
   {|\psi _{\text{W}}\rangle } &\!\! { \equiv\rule{0pt}{8.5pt} {\textstyle{1 \over {\sqrt 4 }}}(|1,\!1,\!1,\!2\rangle \! +\! |1,\!1,\!2,\!1\rangle  \!+\! |1,\!2,\!1,\!1\rangle  \!+\! |2,\!1,\!1,\!1\rangle )}  \\
   {|\psi _{\text{Rand}} \rangle } &\!\! { \equiv\rule{0pt}{8.5pt} \sum\nolimits_{k = 1}^{16} {a_k |k\rangle } ;\;\;\sum\nolimits_{k = 1}^{16} {|a_k |^2 }  = 1},  \\
\end{array}
\label{eq:14}
\end{equation}
where \smash{$|\Phi _{\text{BP}} \rangle \! \equiv\! |\Phi ^ +  \rangle  \otimes |\Phi ^ +  \rangle$} is the Bell product from \Eq{4}, $|\psi _{\text{W}}\rangle$ is the W state \cite{DuVC}, and $|\psi _{\text{Rand}} \rangle$ is a random $16$-level pure state, where here and throughout we use basis abbreviation $\{|1\rangle,|2\rangle,\ldots,|n\rangle\}\equiv\{|1,\!1,\!1,\!1\rangle,$ $|1,\!1,\!1,\!2\rangle,\ldots,|2,\!2,\!2,\!2\rangle\}$.  The states $|\Phi _{\text{GHZ}} \rangle$, $|\Phi _{\text{BP}} \rangle$, and $|\Phi _{\text{F}} \rangle$ are taken from the set of maximally $\text{F}_N$-entangled true-generalized X (TGX) states (see \App{F}, \hyperlink{HedE}{\cite*[App.{\kern 2.5pt}D]{HedE}}, and \cite{HedX}), chosen from the subset including $|1\rangle$, generated by the 13-step algorithm of \cite{HedE}.  Thus, \Eq{14} provides three maximally $\text{F}_N$-entangled states and two other kinds of states for comparison. 

The subset \smash{$\{|\Phi _{\text{GHZ}} \rangle,|\Phi _{\text{BP} } \rangle,|\Phi _{\text{F}} \rangle\}$} was chosen from the full set of $\text{F}_N$-entangled TGX states since they produced distinct results for the measures under test and included $|\Phi _{\text{GHZ}} \rangle$.  See \App{G} for the full set initially used. 

As an example showing the partitional ents involved, unnormalized expansion of the FGM ent in \Eq{6} is
\begin{equation}
\widetilde{\Upsilon}_{\text{FGM}} (\rho ) = \Upsilon _{\text{GM}_2 }  + \Upsilon _{\text{GM}_3 }  + \Upsilon _{\text{GM}_N },
\label{eq:15}
\end{equation}
where $\Upsilon _{\text{GM}_2 } \! =\!\min \{ \Upsilon ^{(1|2,3,4)} {\kern -0.5pt},\!\Upsilon ^{(2|1,3,4)}{\kern -0.5pt} {\kern -0.5pt},\!\Upsilon ^{(3|1,2,4)} {\kern -0.5pt},\!\Upsilon ^{(4|1,2,3)},$ $\Upsilon ^{(1,2|3,4)} {\kern -0.5pt},\!\Upsilon ^{(1,3|2,4)} {\kern -0.5pt},\!\Upsilon ^{(1,4|2,3)}{\kern -0.5pt} \}$,{\kern -1.8pt} and{\kern -1.8pt} $\Upsilon _{\text{GM}_3 } \! =\! \min \{{\kern -0.5pt} \Upsilon ^{(1|2|3,4)} {\kern -0.5pt},\!{\kern -0.8pt}$ \mbox{$\Upsilon ^{(1|3|2,4)} {\kern -0.5pt},\!{\kern -0.8pt}\Upsilon ^{(1|4|2,3)} {\kern -0.5pt},\!{\kern -0.8pt}\Upsilon ^{(2|3|1,4)} {\kern -0.5pt},\!{\kern -0.8pt}\Upsilon ^{(2|4|1,3)} {\kern -0.5pt},\!{\kern -0.8pt}\Upsilon ^{(3|4|1,2)} {\kern -0.8pt}\}$,{\kern -2.5pt} and{\kern -1.5pt} $\Upsilon _{\!\text{GM}_N }$} $=\!\min \{ \Upsilon ^{(1|2|3|4)} \} \! =\! \Upsilon$, all using $N$-mode partitional ents.
\subsection{\label{sec:III.A}Tests and Analysis of FGM Ent}
\begin{figure}[H]
\centering
\includegraphics[width=1.00\linewidth]{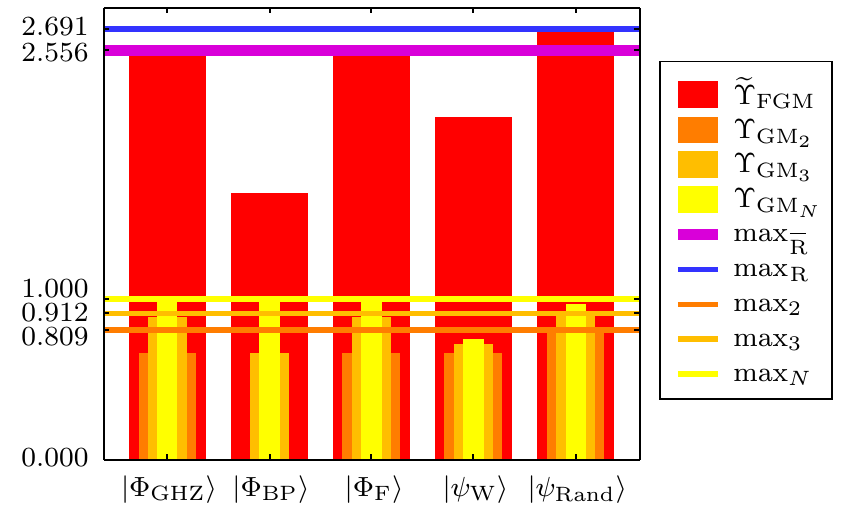}%
\vspace{-6pt}
\caption[]{(color online) Unnormalized FGM ent \smash{$\widetilde{\Upsilon}_{\text{FGM}}$} of \Eq{6} for the test states of \Eq{14}, with only the \smash{$|\psi _{\text{Rand}} \rangle$} that maximizes it after $1000$ random pure states were tested. The normalized \smash{$\text{GM}_{k}$} ents \smash{$\Upsilon_{\text{GM}_k}$} of \Eq{7} are also shown, and the maximum of each over all test states is \smash{$\max_{k}\equiv \max(\Upsilon_{\text{GM}_k})$}, while \smash{$\max_{{\kern 0.5pt}\overline{\rule{0pt}{5pt}{\kern -0.1pt}\text{R}{\kern -0.6pt}}}\equiv$} \smash{$\max(\widetilde{\Upsilon}_{\text{FGM}})$} applies to all nonrandom test states, and \smash{$\max_{{\kern 0.5pt}\text{R}}\equiv\max(\widetilde{\Upsilon}_{\text{FGM}}(\rho_{|\psi _{\text{Rand}} \rangle}))$}.}
\label{fig:3}
\end{figure}

\Figure{3} explores the FGM ent of the test states in \Eq{14}, with only the results for the \smash{$|\psi _{\text{Rand}} \rangle$} that had the highest unnormalized FGM ent \smash{$\widetilde{\Upsilon}_{\text{FGM}}$} out of $1000$ random states.
As \Fig{3} shows, the maximally $\text{F}_N$-entangled ($N=4$) states \smash{$|\Phi _{\text{GHZ}} \rangle$} and \smash{$|\Phi _{\text{F}} \rangle$} have identical results for all \smash{$\Upsilon_{\text{GM}_k}$} and \smash{$\widetilde{\Upsilon}_{\text{FGM}}$}, but the maximally $\text{F}_N$-entangled \smash{$|\Phi _{\text{BP}} \rangle$} has \smash{$\Upsilon_{\text{GM}_2}=0$} and therefore a lower \smash{$\widetilde{\Upsilon}_{\text{FGM}}$}, despite matching \smash{$|\Phi _{\text{GHZ}} \rangle$} and \smash{$|\Phi _{\text{F}} \rangle$} for \smash{$\Upsilon_{\text{GM}_N}$}.  As expected, \smash{$|\psi _{\text{W}} \rangle$} is not maximal in any quantity, but still has fairly high values, and actually \textit{outperforms} \smash{$|\Phi _{\text{BP}} \rangle$} in \smash{$\Upsilon_{\text{GM}_2}$}, \smash{$\Upsilon_{\text{GM}_3}$}, and \smash{$\widetilde{\Upsilon}_{\text{FGM}}$}, despite its lower \smash{$\Upsilon_{\text{GM}_N}$}.

Interestingly, \smash{$|\psi _{\text{Rand}} \rangle$} reached a \textit{higher} \smash{$\widetilde{\Upsilon}_{\text{FGM}}$} than all other test states, since its \smash{$\Upsilon_{\text{GM}_2}$} and \smash{$\Upsilon_{\text{GM}_3}$} are \textit{higher} than those of the other states, while its \smash{$\Upsilon_{\text{GM}_N}$} is still slightly \textit{lower} than $1$.  This proves by example that there are nonmaximally-$\text{F}_N$-entangled FGM-entangled states with higher \smash{$\widetilde{\Upsilon}_{\text{FGM}}$} than maximally $\text{F}_N$-entangled states.
\subsection{\label{sec:III.B}Tests and Analysis of FSM Ent}
\begin{figure}[H]
\centering
\includegraphics[width=1.00\linewidth]{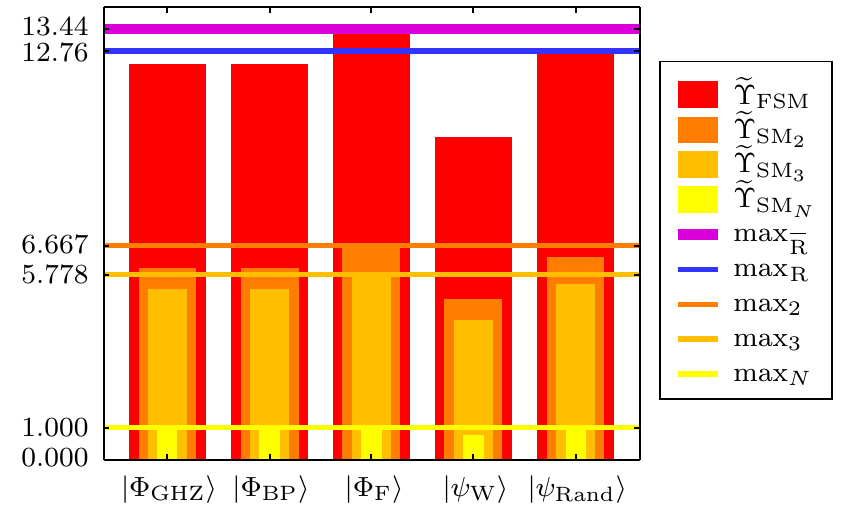}%
\vspace{-6pt}
\caption[]{(color online) Unnormalized FSM ent \smash{$\widetilde{\Upsilon}_{\text{FSM}}$} of \Eq{8} for the test states of \Eq{14}, with only the \smash{$|\psi _{\text{Rand}} \rangle$} that maximizes it over $1000$ random pure states. The \textit{unnormalized} \smash{$\text{SM}_{k}$} ents \smash{$\widetilde{\Upsilon}_{\text{SM}_k}$} of \Eq{9} are also shown, and the maximum of each over all test states is \smash{$\max_{k}\equiv \max(\widetilde{\Upsilon}_{\text{SM}_k})$}, while \smash{$\max_{{\kern 0.5pt}\overline{\rule{0pt}{5pt}{\kern -0.1pt}\text{R}{\kern -0.6pt}}}\equiv$} \smash{$\max(\widetilde{\Upsilon}_{\text{FSM}})$} applies to all nonrandom test states, and \smash{$\max_{{\kern 0.5pt}\text{R}}\equiv\max(\widetilde{\Upsilon}_{\text{FSM}}(\rho_{|\psi _{\text{Rand}} \rangle}))$}.}
\label{fig:4}
\end{figure}

Here, \Fig{4} applies the same tests as in \Fig{3}, but this time for the FSM ent of \Eq{8}.
Here, we see that \smash{$|\Phi _{\text{BP}} \rangle$} has the \textit{same} \smash{$\widetilde{\Upsilon}_{\text{SM}_k}$} and \smash{$\widetilde{\Upsilon}_{\text{FSM}}$} as \smash{$|\Phi _{\text{GHZ}} \rangle$}, but that \textit{both} \smash{$|\Phi _{\text{GHZ}} \rangle$} and \smash{$|\Phi _{\text{BP}} \rangle$} \textit{underperform} \smash{$|\Phi _{\text{F}} \rangle$} in terms of \smash{$\widetilde{\Upsilon}_{\text{SM}_2}$}, \smash{$\widetilde{\Upsilon}_{\text{SM}_3}$}, and \smash{$\widetilde{\Upsilon}_{\text{FSM}}$}, despite all three states being maximally $\text{F}_N$-entangled.  This time, \smash{$|\psi _{\text{W}} \rangle$} underperforms all other test states in every area, while \smash{$|\psi _{\text{Rand}} \rangle$} \textit{outperforms} \smash{$|\Phi _{\text{GHZ}} \rangle$} and \smash{$|\Phi _{\text{BP}} \rangle$}, while still underperforming \smash{$|\Phi _{\text{F}} \rangle$}, suggesting that \smash{$|\Phi _{\text{F}} \rangle$} may be maximal in all quantities being measured.

Thus, this proves by example that some maximally $\text{F}_N$-entangled states are more FSM-entangled than others, even \smash{$|\Phi _{\text{GHZ}} \rangle$}, and \textit{suggests} that maximally FSM-entangled states may also be maximal for all \smash{$\Upsilon_{\text{SM}_{k}}$}.
\subsection{\label{sec:III.C}Tests and Analysis of FDM Ent}
Here we apply the same tests as in \Sec{III.A} and \Sec{III.B}, this time to the FDM ent of \Eq{10}.
\begin{figure}[H]
\centering
\includegraphics[width=1.00\linewidth]{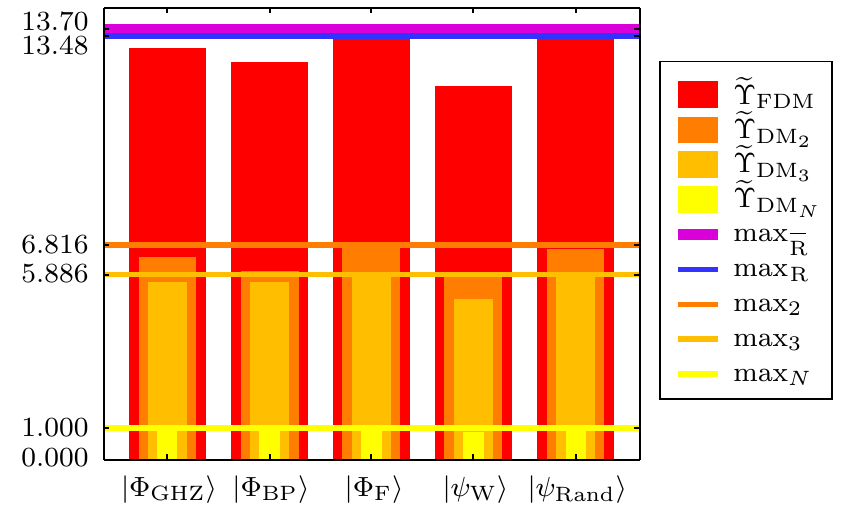}%
\vspace{-6pt}
\caption[]{(color online) Unnormalized FDM ent \smash{$\widetilde{\Upsilon}_{\text{FDM}}$} (or \textit{ent-concurrence}) of \Eq{10} for the test states of \Eq{14}, with only the \smash{$|\psi _{\text{Rand}} \rangle$} that maximizes it over $1000$ random pure states. The \textit{unnormalized} \smash{$\text{DM}_{k}$} ents \smash{$\widetilde{\Upsilon}_{\text{DM}_k}$}  of \Eq{11} are also shown, and the maximum of each over all test states is \smash{$\max_{k}\equiv \max(\widetilde{\Upsilon}_{\text{DM}_k})$}, while \smash{$\max_{{\kern 0.5pt}\overline{\rule{0pt}{5.3pt}{\kern -0.1pt}\text{R}{\kern -0.6pt}}}\equiv\max(\widetilde{\Upsilon}_{\text{FDM}})$} applies to all nonrandom test states, and \smash{$\max_{{\kern 0.5pt}\text{R}}\!\equiv\!\max(\widetilde{\Upsilon}_{\text{FDM}}(\rho_{|\psi _{\text{Rand}} \rangle}))$}.}
\label{fig:5}
\end{figure}
\Figure{5} shows \textit{different} results for \smash{$|\Phi _{\text{GHZ}} \rangle$} and \smash{$|\Phi _{\text{BP}} \rangle$}, which is mainly due to the one biseparability of \smash{$|\Phi _{\text{BP}} \rangle$}, making the FDM ent the only measure of these three that distinguishes between the separability of these states, and yet does not neglect the other bipartite entanglement correlations in \smash{$|\Phi _{\text{BP}} \rangle$}.  Again, \smash{$|\Phi _{\text{F}} \rangle$} seems to outperform all other states in every measure, as suggested by the fact that \smash{$|\psi _{\text{Rand}} \rangle$} seems to approach its performance, and \smash{$|\psi _{\text{W}} \rangle$} slightly underperforms in every area.
\subsection{\label{sec:III.D}Comparison of FGM, FSM, and FDM Ents}
The most important difference between \smash{$\Upsilon_{\text{FGM}}$} and both \smash{$\Upsilon_{\text{FSM}}$} and \smash{$\Upsilon_{\text{FDM}}$} is that \smash{${\Upsilon}_{\text{GM}_2}(\rho_{|\Phi _{\text{BP}} \rangle})=0$} while \smash{$\widetilde{\Upsilon}_{\text{SM}_2}(\rho_{|\Phi _{\text{BP}} \rangle})\neq 0$} and \smash{$\widetilde{\Upsilon}_{\text{DM}_2}(\rho_{|\Phi _{\text{BP}} \rangle})\neq 0$}. The fact that not all bipartitions of \smash{$|\Phi _{\text{BP}} \rangle$} are separable means that there are \textit{some} bipartitions that have \textit{entanglement correlations}, and the \textit{sum} over all $2$-partitional quantities in \smash{$\widetilde{\Upsilon}_{\text{SM}_2}$} and \smash{$\widetilde{\Upsilon}_{\text{DM}_2}$} is why they detect these correlations, while the \textit{minimum} over all $2$-partitions in \smash{${\Upsilon}_{\text{GM}_2}$} is why it \textit{misses those entanglement correlations}, reporting ``zero.''  Therefore \smash{${\Upsilon}_{\text{GM}_2}$} \textit{is not sufficient to detect all entanglement correlations of $2$-partitions of a state}, so \smash{$\Upsilon_{\text{FGM}}$} is not sufficient to detect all entanglement correlations.

Thus, we must make an important new distinction; \textit{presence of separability in a particular $k$-partition is not sufficient to claim absence of entanglement correlations over all $k$-partitions for a given $k$.}  Since \textit{nonseparable nonlocal correlations} (NSNLC) are what the word \textit{entanglement} really means, we must require \textit{the necessary and sufficient detection of the presence of any NSNLC} as our prime criterion for what constitutes an entanglement measure.  We state all of this in the following theorem.

\hypertarget{Thm:TrueEnt}{\textbf{True-Entanglement Theorem:~~}}The\hsp{-0.6} \textit{absence\hsp{-0.6} of\hsp{-0.6} separability\hsp{-0.6} between\hsp{-0.6} any\hsp{-0.6} partitions}\hsp{-0.6} is\hsp{-0.6} necessary\hsp{-0.6} and\hsp{-0.6} sufficient\hsp{-0.6} for\hsp{-0.6} the\hsp{-0.6} \textit{presence\hsp{-0.6} of\hsp{-0.6} any\hsp{-0.6} entanglement\hsp{-0.6} correlations},\hsp{-0.6} and\hsp{-0.6} therefore\hsp{-0.5} the\hsp{-0.5} presence\hsp{-0.5} of\hsp{-0.5} true\hsp{-0.5} entanglement.\hsp{-0.5} In\hsp{-0.5} other\hsp{-0.5} words;\hsp{-0.5} the\hsp{-0.5} \textit{presence\hsp{-0.5} of\hsp{-0.5} separability\hsp{-0.5} between\hsp{-0.5} all\hsp{-0.5} partitions}\hsp{-0.5} is\hsp{-0.5} necessary\hsp{-0.5} and\hsp{-0.5} sufficient\hsp{-0.5} for\hsp{-0.5} the\hsp{-0.5} \textit{absence\hsp{-0.5} of\hsp{-0.5} all\hsp{-0.5} entanglement\hsp{-0.5} correlations},\hsp{-0.5} and\hsp{-0.4} thus\hsp{-0.4} the\hsp{-0.4} absence\hsp{-0.4} of\hsp{-0.4} true\hsp{-0.4} entanglement.

This\hsp{-0.2} theorem\hsp{-0.2} reflects\hsp{-0.2} the\hsp{-0.2} fact\hsp{-0.2} that\hsp{-0.2} although\hsp{-0.2} a\hsp{-0.2} state\hsp{-0.2} may\hsp{-0.2} be\hsp{-0.2} separable\hsp{-0.2} over\hsp{-0.2} one\hsp{-0.2} particular\hsp{-0.2} $k$-partition,\hsp{-0.2} that\hsp{-0.2} is\hsp{-0.2} not\hsp{-0.2} sufficient\hsp{-0.2} to\hsp{-0.2} conclude\hsp{-0.2} the\hsp{-0.2} absence\hsp{-0.2} of\hsp{-0.2} all\hsp{-0.2} $k$-partite\hsp{-0.2} entanglement\hsp{-0.2} correlations,\hsp{-0.2} so\hsp{-0.2} sub-$N$\hsp{-0.2} $k$-separability\hsp{-0.2} is\hsp{-0.2} not\hsp{-0.2} a\hsp{-0.2} sufficient\hsp{-0.2} criterion\hsp{-0.2} for\hsp{-0.2} the\hsp{-0.2} absence\hsp{-0.2} of\hsp{-0.2} all\hsp{-0.2} $k$-partite\hsp{-0.2} NSNLC.

The likely reason that $\text{GM}_k$-entanglement was defined with the min function is that the presence of separability in any one bipartition \textit{is} sufficient to claim the absence of entanglement correlations for \textit{bipartite} systems, since there is \textit{only one bipartition}.  So if separability were our only concern for multimode systems, then $\text{GM}_k$-entanglement would be correctly defined because if a state is in \textit{any way $k$-separable}, then its $\text{GM}_k$-entanglement is $0$. (After all, GM measures \textit{do} correctly indicate whether a pure state is a product over at least one $k$-partition.)  But since we have just seen examples that $k$-separability is not sufficient to conclude the absence of all $k$-partite NSNLC, and since NSNLC are what entanglement \textit{really is}, then $\text{GM}_k$-``entanglement'' is \textit{not} really a measure of $k$-partite entanglement.

Unfortunately, there is now quite a lot of literature that uses the terminology (GM)-``$k$-entangled'' to describe a condition that is insufficient to determine the presence of entanglement correlations over all $k$-partitions. Our\hsp{-0.5} remedy\hsp{-0.5} for\hsp{-0.5} this\hsp{-0.5} was\hsp{-0.5} to\hsp{-0.5} observe\hsp{-0.5} the\hsp{-0.5} fact\hsp{-0.5} that\hsp{-0.5} the\hsp{-0.5} absence\hsp{-0.5} of $k$-separability is the condition of all $k$-partitions not having $k$-mode product form, which motivated us to sum the $N$-mode $k$-partitional entanglement values over all $k$-partitions in our various candidate entanglement measures in \Sec{II}. These sums gave us candidate $\text{F}_k$-entanglement values which were further added over all $k$ values to construct candidate measures for detecting all possible entanglement correlations in a given pure state.

However, the \hyperlink{Thm:TrueEnt}{True-Entanglement Theorem} alone is not sufficient to \textit{distinguish between} states like \smash{$|\Phi _{\text{GHZ}} \rangle$} and \smash{$|\Phi _{\text{BP}} \rangle$}.  One missing concept is that the states that are most entangled have the most \textit{simultaneous} entanglement correlations over all possible partitions. Since this is exactly what the FSM ent measures, then it is both necessary and sufficient to \textit{detect} all entanglement correlations, \textit{and} it measures their simultaneous presence, providing an \textit{ordering} for multipartite entangled states.  

Yet the FSM ent's ordering is \textit{still} not sufficient to reveal the difference between states like \smash{$|\Phi _{\text{GHZ}} \rangle$} and \smash{$|\Phi _{\text{BP}} \rangle$}, as seen in \Fig{4}. Therefore, by also requiring that the simultaneous entanglement correlations be \textit{as evenly and as widely distributed as possible}, we attain ordering criteria that distinguish \smash{$|\Phi _{\text{GHZ}} \rangle$} and \smash{$|\Phi _{\text{BP}} \rangle$} without sacrificing information about entanglement correlations. The FDM ent (ent-concurrence) may achieve this, as seen in \Fig{5}.

However,\hsp{-0.5} the\hsp{-0.5} \hyperlink{Thm:TrueEnt}{True-Entanglement\hsp{-0.5} Theorem}\hsp{-0.5} \textit{only applies to applications of entanglement in the full Hilbert space of $\rho$}. For applications of entanglement in the \textit{reductions}, other principles are involved, as discussed in \Sec{VII}.
\section{\label{sec:IV}Candidate Mixed-State Measures}
\begin{figure}[H]
\centering
\vspace{-12pt}
\setlength{\unitlength}{0.01\linewidth}
\begin{picture}(100,0)
\put(1,24){\hypertarget{Sec:IV}{}}
\end{picture}
\end{figure}
\vspace{-39pt}
Here we list the mixed-state entanglement-measure candidates that we test in \Sec{V}. In all cases, $\hat{E}(\rho)$ means the convex-roof extension of a pure-state measure $E(\rho)$ to handle mixed-state input (see \hyperlink{HedE}{\cite*[App.{\kern 2.5pt}J]{HedE}}).
\begin{itemize}[leftmargin=*,labelindent=4pt]\setlength\itemsep{0pt}
\item[\textbf{1.}]\hypertarget{mixedmeasure:1}{}The FGM ent of formation (using $\Upsilon_{\text{FGM}} (\rho )$ from \Eq{6}):
\begin{equation}
\hat{\Upsilon}_{\text{FGM}} (\rho ) \equiv\! \mathop {\min }\limits_{\{ p_j ,\rho _j \} |_{\rho  = \sum\nolimits_j {p_j \rho _j } } }\!\! \left( {\sum\nolimits_j {p_j \Upsilon _{\text{FGM}} (\rho _j )} }\! \right)\!.
\label{eq:16}
\end{equation}
\item[\textbf{2.}]\hypertarget{mixedmeasure:2}{}The strict FGM (SFGM) ent of formation:
\begin{equation}
\Upsilon _{\text{SFGM}} (\rho ) \equiv \frac{1}{M_{\text{SFGM}}}\sum\limits_{k = 2}^N {\Upsilon _{\text{SGM}_k } (\rho )},
\label{eq:17}
\end{equation}
where \smash{$M_{\text{SFGM}}$} is a normalization factor, and the strict \smash{$\text{GM}_k$} (\smash{$\text{SGM}_k$}) ent of formation is
\begin{equation}
\Upsilon _{\text{SGM}_k } (\rho ) \equiv \min (\{ \hat{\Upsilon}^{(\mathbf{N}_h^{(\mathbf{k})} )} (\rho )\} ),
\label{eq:18}
\end{equation}
where \smash{$\{ \hat{\Upsilon}^{(\mathbf{N}_h^{(\mathbf{k})} )}(\rho ) \}$} is the set of \textit{convex-roof extensions} of all $N$-mode $k$-partitional ents. Here, the minimum \textit{over all convex-roof-extensions} of a given kind of $k$-partitional ent ensures that if a state achieves \textit{strict $k$-separability}, \textit{all} of its optimal-decomposition pure states are $k$-separable \textit{over the same partitions}.
\item[\textbf{3.}]\hypertarget{mixedmeasure:3}{}The FSM ent of formation (using \smash{$\Upsilon _{\text{FSM}} (\rho)$} from \Eq{8}):
\begin{equation}
\hat{\Upsilon}_{\text{FSM}} (\rho ) \equiv\! \mathop {\min }\limits_{\{ p_j ,\rho _j \} |_{\rho  = \sum\nolimits_j {p_j \rho _j } } } \!\!\left( {\sum\nolimits_j {p_j \Upsilon _{\text{FSM}} (\rho _j )} }\! \right)\!.
\label{eq:19}
\end{equation}
\item[\textbf{4.}]\hypertarget{mixedmeasure:4}{}The FDM ent of formation (with \smash{$\Upsilon _{\text{FDM}} (\rho)$} from \Eq{10}):
\begin{equation}
\hat{\Upsilon}_{\text{FDM}} (\rho ) \equiv\! \mathop {\min }\limits_{\{ p_j ,\rho _j \} |_{\rho  = \sum\nolimits_j {p_j \rho _j } } } \!\!\left( {\sum\nolimits_j {p_j \Upsilon _{\text{FDM}} (\rho _j )} }\! \right)\!.
\label{eq:20}
\end{equation}
\end{itemize}
We do not use a ``hat'' if a convex-roof extension (CRE) has been applied already within the function.
\section{\label{sec:V}Tests of Candidate Mixed-State Measures}
\begin{figure}[H]
\centering
\vspace{-12pt}
\setlength{\unitlength}{0.01\linewidth}
\begin{picture}(100,0)
\put(1,27){\hypertarget{Sec:V}{}}
\end{picture}
\end{figure}
\vspace{-39pt}
Limiting ourselves to rank-2 mixed states (since CREs of those are practical to approximate) we use test states,
\begin{equation}
\begin{array}{*{20}l}
   {\rho _{\text{GHZ} + 1} } &\!\! { \equiv \frac{1}{2}(|\Phi _{\text{GHZ}} \rangle \langle \Phi _{\text{GHZ}} | + |1,\!1,\!1,\!1\rangle \langle 1,\!1,\!1,\!1|)}  \\
   {\rho _{2\text{-sep}} } &\!\!\rule{0pt}{9.8pt} { \equiv \frac{1}{2}(\rho _{|\Phi ^ +  \rangle  \otimes |\Phi ^ +  \rangle }  + \rho_{|1\rangle}^{(1)} \otimes |\Phi _{\text{GHZ}_3 } \rangle \langle \Phi _{\text{GHZ}_3 } |)}  \\
   {\rho _{\text{F} + 1} } &\!\!\rule{0pt}{9.8pt} { \equiv \frac{1}{2}(|\Phi _{\text{F}} \rangle \langle \Phi _{\text{F}} | + |1,\!1,\!1,\!1\rangle \langle 1,\!1,\!1,\!1|)}  \\
   {\rho _{\text{MME} } } &\!\!\rule{0pt}{9.8pt} { \equiv \frac{1}{2}(\rho _{{\textstyle{1 \over {\sqrt 2 }}}(|1,{\kern -0.5pt}1,{\kern -0.5pt}1,{\kern -0.5pt}1\rangle  + |2,{\kern -0.5pt}2,{\kern -0.5pt}2,{\kern -0.5pt}1\rangle )}  \!+\! \rho _{{\textstyle{1 \over {\sqrt 2 }}}(|1,{\kern -0.5pt}1,{\kern -0.5pt}1,{\kern -0.5pt}2\rangle  + |2,{\kern -0.5pt}2,{\kern -0.5pt}2,{\kern -0.5pt}2\rangle )} )}  \\
   {\rho _{|\Phi_{\text{F}}  \rangle } } &\!\!\rule{0pt}{9.8pt} { \equiv |\Phi _{\text{F}} \rangle \langle \Phi _{\text{F}} |,}  \\
\end{array}
\label{eq:21}
\end{equation}
where \smash{$\rho _{\shiftmath{0.5pt}{|A\rangle} }  \!\equiv\! |A\rangle \langle A|$}, and \smash{$|\Phi _{\text{GHZ}} \rangle$}, \smash{$|\Phi ^ +  \rangle  \otimes |\Phi ^ +  \rangle  \!\equiv\! |\Phi _{\text{BP}} \rangle$}, and \smash{$|\Phi _{\text{F}} \rangle$} are from \Eq{14}, \smash{$\rho_{\shiftmath{0.5pt}{\scalemath{0.9}{|1\rangle}}}^{{\kern 0.5pt}\shiftmath{0.5pt}{\scalemath{0.9}{(1)}}}$} is the first computational basis state for the mode-$1$ qubit, and \smash{$|\Phi _{\text{GHZ}_3 } \rangle  \equiv$} \smash{$ {\textstyle{1 \over {\sqrt 2 }}}(|1,\!1,\!1\rangle \langle 1,\!1,\!1|\! +\! |2,\!2,\!2\rangle \langle 2,\!2,\!2|)$}{\kern -0.5pt} is{\kern -0.5pt} a{\kern -0.5pt} $3$-qubit{\kern -0.5pt} GHZ{\kern -0.5pt} state.

We{\kern -1pt} include{\kern -1pt} \smash{$\rho _{\text{GHZ} + 1}$}{\kern -1pt} and{\kern -1pt} \smash{$\rho _{\text{F} + 1}$}{\kern -1pt} because{\kern -1pt} they{\kern -1pt} are{\kern -1pt} mixtures of highly entanglement-correlated states with a basis state they already include, to see how the candidate measures rate this lowering of entanglement correlation.

To test a state like \Eq{2}, \smash{$\rho _{2\text{-sep}}$} decomposes into pure states that are each $2$-separable \textit{in different ways}, where each part has strong internal entanglement correlations.

The state \smash{$\rho _{\text{MME} }$}, when viewed as a $2 \times 8$ system is a \textit{mixed state with the same entanglement as a maximally entangled pure state}. Rediscovered in the present work, this phenomenon was originally discovered in \cite{CaBC,LZFF}, and called ``mixed maximally entangled (MME) states.''  It{\kern -1pt} is{\kern -1pt} easy{\kern -1pt} to{\kern -1pt} prove{\kern -1pt} that{\kern -1pt} \textit{all}{\kern -1pt} decompositions{\kern -1pt} of{\kern -1pt} such{\kern -1pt} states{\kern -1pt} consist{\kern -1pt} of{\kern -1pt} maximally{\kern -1pt} entangled{\kern -1pt} pure{\kern -1pt} decomposition{\kern -1pt} states, yielding an entanglement of $1$ by any unit-normalized convex-roof (or nearest-separable-state \cite{StKB}) measure.

We use \textit{pure} state $\rho _{|\Phi_{\text{F}}  \rangle }$ as a reference since it had near-highest values in the GM measures of \Fig{3}, and it may have the highest values for the SM measures in \Fig{4} and the DM measures in \Fig{5}.
\subsection{\label{sec:V.A}Tests and Analysis of Mixed-Input FGM Ents}
\begin{figure}[H]
\centering
\includegraphics[width=1.00\linewidth]{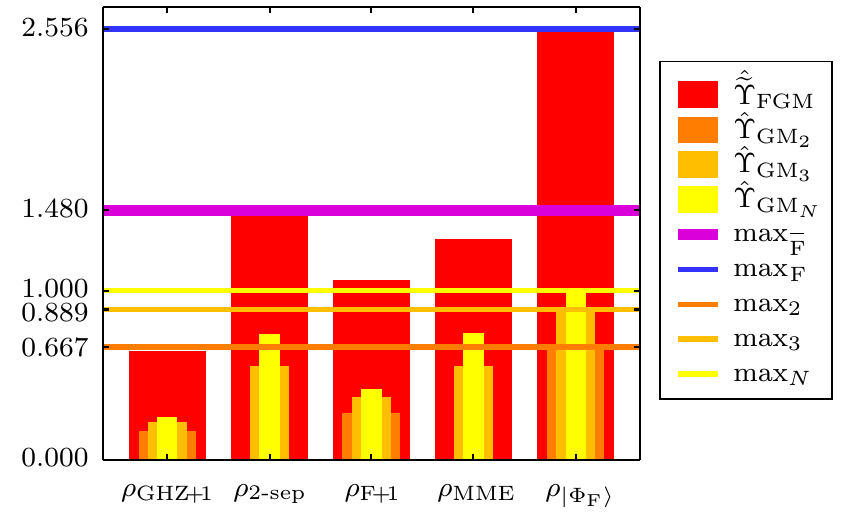}%
\vspace{-6pt}
\caption[]{(color online) Unnormalized FGM ent of formation \smash{$\scalemath{0.80}{\hat{\widetilde{\Upsilon}}}_{\text{FGM}}$} of \Eq{16} approximated for the test states of \Eq{21}. The (normalized) \smash{$\text{GM}_{k}$} ents of formation \smash{${\hat{\Upsilon}}_{\text{GM}_{k}}$} (CREs of \Eq{7}) are also approximated, and the maximum of each over all test states is \smash{$\max_{k}\equiv \max(\hat{\Upsilon}_{\text{GM}_k})$}, while \smash{$\max_{{\kern 0.5pt}\overline{\rule{0pt}{5.3pt}{\kern -0.1pt}\text{F}{\kern -0.6pt}}}\equiv$} \smash{$\max(\scalemath{0.80}{\hat{\widetilde{\Upsilon}}}_{\text{FGM}})$} applies to all non$\rho _{|\Phi_{\text{F}}\rangle}$ test states, and \smash{$\max_{{\kern 0.5pt}\text{F}}\equiv\max(\scalemath{0.80}{\hat{\widetilde{\Upsilon}}}_{\text{FGM}}(\rho_{|\Phi _{\text{F}} \rangle}))$}. The CRE approximations used $900$ decompositions, as in \cite{HedE}.}
\label{fig:6}
\end{figure}

\Figures{6}{}{7}{} show similar results but differ in subtle ways briefly explained after each.
The main items of interest in \Fig{6} are the fact that both \smash{$\rho _{2\text{-sep}}$} and \smash{$\rho _{\text{MME} }$} have \smash{${\hat\Upsilon}_{\text{GM}_2}=0$}, and while we expect this to be true from the way the FGM ent minimizes over all bipartitions for each decomposition state within the larger minimization of the CRE, it shows that GM measures \textit{ignore} entanglement correlations, since \smash{$\rho _{\text{MME} }$} in particular has the maximal entanglement of a pure state for the $(1|2,3,4)$ bipartition, as mentioned earlier.
\begin{figure}[H]
\centering
\includegraphics[width=1.00\linewidth]{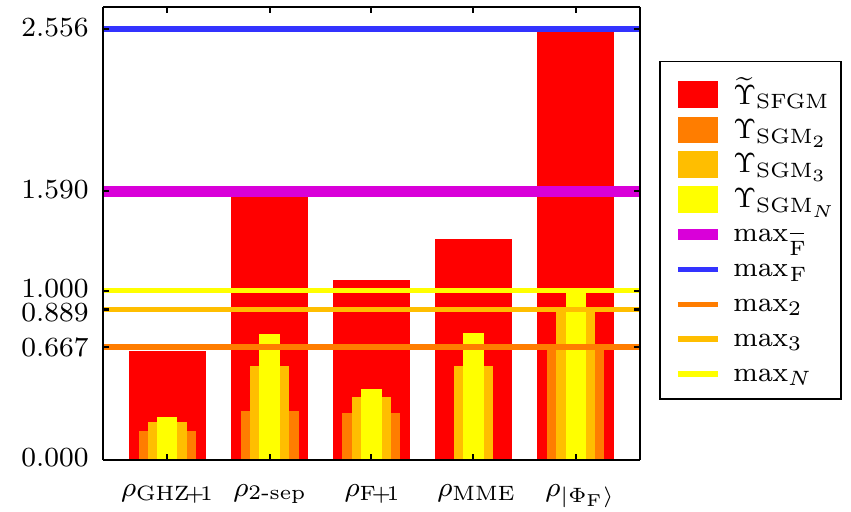}%
\vspace{-6pt}
\caption[]{(color online) Unnormalized SFGM ent of formation \smash{${\widetilde{\Upsilon}}_{\text{SFGM}}$} of \Eq{17} approximated for the test states of \Eq{21}. The (normalized) \smash{$\text{SGM}_{k}$} ents of formation \smash{${{\Upsilon}}_{\text{SGM}_{k}}$} of \Eq{18} are also approximated, and the maximum of each over all test states is \smash{$\max_{k}\equiv \max({\Upsilon}_{\text{SGM}_k})$}, while \smash{$\max_{{\kern 0.5pt}\overline{\rule{0pt}{5.3pt}{\kern -0.1pt}\text{F}{\kern -0.6pt}}}\equiv$} \smash{$\max({\widetilde{\Upsilon}}_{\text{SFGM}})$} applies to all non$\rho _{|\Phi_{\text{F}}\rangle}$ test states, and \smash{$\max_{{\kern 0.5pt}\text{F}}\equiv\max({\widetilde{\Upsilon}}_{\text{SFGM}}(\rho_{|\Phi _{\text{F}} \rangle}))$}. CRE approximations used $900$ decompositions.}
\label{fig:7}
\end{figure}
The \textit{strict} version, the SFGM ent of formation \smash{${\widetilde{\Upsilon}}_{\text{SFGM}}$} in \Fig{7} does slightly better than \smash{$\scalemath{0.80}{\hat{\widetilde{\Upsilon}}}_{\text{FGM}}$} in \Fig{6} because it correctly detects that no bipartitions of \smash{$\rho _{2\text{-sep}}$} are without entanglement correlation since its \smash{${\Upsilon}_{\text{SGM}_2}\neq 0$}, but it still \textit{completely ignores} the maximal bipartite entanglement correlation in \smash{$\rho _{\text{MME} }$}, for which \smash{${\Upsilon}_{\text{SGM}_2}= 0$}.

We discuss further issues with GM measures regarding states like \smash{$\rho _{2\text{-sep}}$} and \Eq{2} in \App{H}.
\subsection{\label{sec:V.B}Tests and Analysis of Mixed-Input FSM Ent}
\begin{figure}[H]
\centering
\includegraphics[width=1.00\linewidth]{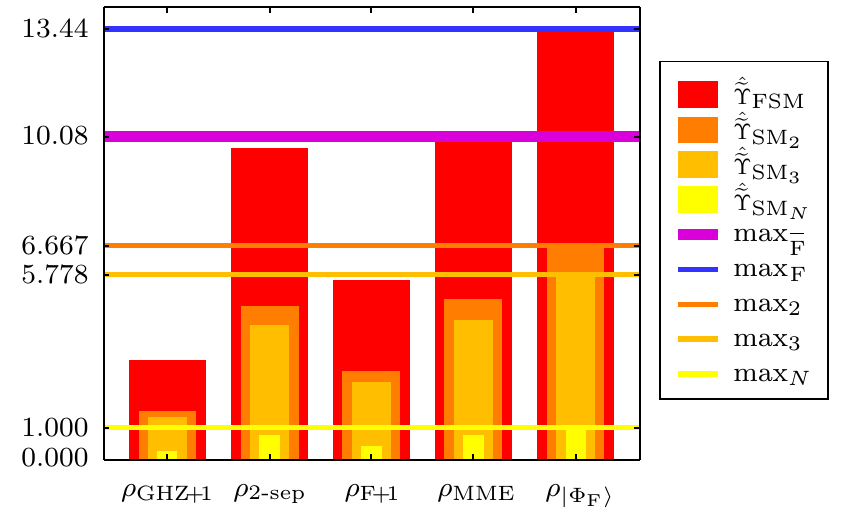}%
\vspace{-6pt}
\caption[]{(color online) Unnormalized FSM ent of formation \smash{\scalebox{0.80}{$\hat{\widetilde{\Upsilon}}$}$_{\text{FSM}}$} of \Eq{19} approximated for the test states of \Eq{21}. The unnormalized \smash{$\text{SM}_{k}$} ents of formation \smash{\scalebox{0.80}{${\hat{\widetilde{\Upsilon}}}$}$_{\text{SM}_{k}}$} (CREs of \Eq{9}) are also approximated, and the maximum of each over all test states is \smash{$\max_{k}\equiv \max($\scalebox{0.80}{$\hat{\widetilde{\Upsilon}}$}$_{\text{SM}_k})$}, while \smash{$\max_{{\kern 0.5pt}\overline{\rule{0pt}{5.3pt}{\kern -0.1pt}\text{F}{\kern -0.6pt}}}\,\equiv\,$} \smash{$\max($\scalebox{0.80}{$\hat{\widetilde{\Upsilon}}$}$_{\text{FSM}})$} applies to all non$\rho _{\shiftmath{0.5pt}{|\Phi_{\text{F}}\rangle}}$ test states, and \smash{$\max_{{\kern 0.5pt}\text{F}}\!\equiv\!\max($\scalebox{0.80}{$\hat{\widetilde{\Upsilon}}$}$_{\text{FSM}}(\rho_{|\Phi _{\text{F}} \rangle}))$}. CRE approximations used $900$ decompositions.}
\label{fig:8}
\end{figure}
\Figure{8} tests unnormalized FSM ent of formation from \Eq{19}, and we see that the known bipartite entanglement correlations in both \smash{$\rho _{2\text{-sep}}$} and \smash{$\rho _{\text{MME} }$} are detected since \smash{\scalebox{0.80}{${\hat{\widetilde{\Upsilon}}}$}$_{\text{SM}_{2}}\neq 0$} for each.
\subsection{\label{sec:V.C}Tests and Analysis of Mixed-Input FDM Ent (The Ent-Concurrence)}
\vspace{-6pt}
\begin{figure}[H]
\centering
\includegraphics[width=1.00\linewidth]{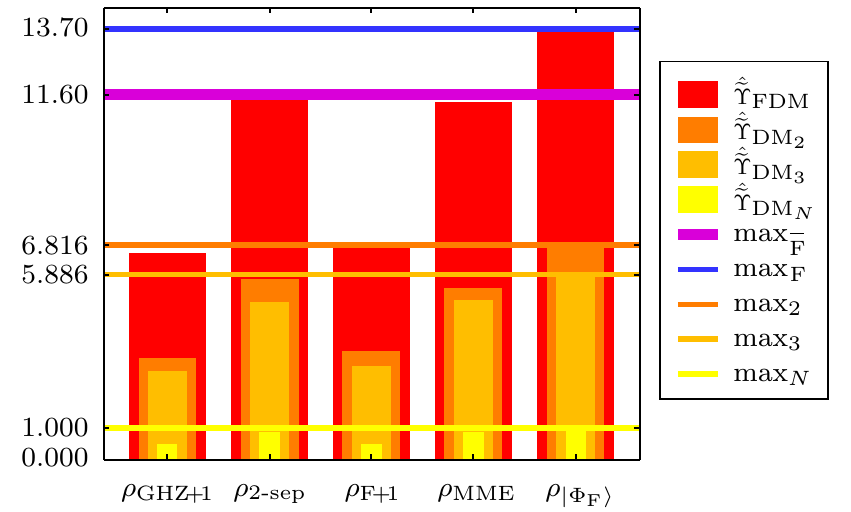}%
\vspace{-6pt}
\caption[]{(color online) Unnormalized FDM ent of formation (ent-concurrence of formation) \smash{\scalebox{0.80}{$\hat{\widetilde{\Upsilon}}$}$_{\text{FDM}}$} of \Eq{20} approximated for the test states of \Eq{21}. The unnormalized \smash{$\text{DM}_{k}$} ents of formation \smash{\scalebox{0.80}{${\hat{\widetilde{\Upsilon}}}$}$_{\text{DM}_{k}}$} (CREs of \Eq{11}) are also approximated, and the maximum of each over all test states is \smash{$\max_{k}\equiv \max($\scalebox{0.80}{$\hat{\widetilde{\Upsilon}}$}$_{\text{DM}_k})$}, while \smash{$\max_{{\kern 0.5pt}\overline{\rule{0pt}{5.3pt}{\kern -0.1pt}\text{F}{\kern -0.6pt}}}\equiv$} \smash{$\max($\scalebox{0.80}{$\hat{\widetilde{\Upsilon}}$}$_{\text{FDM}})$} applies to all non$\rho _{\shiftmath{0.5pt}{|\Phi_{\text{F}}\rangle}}$ test states, and \smash{$\max_{{\kern 0.5pt}\text{F}}\equiv\max($\scalebox{0.80}{$\hat{\widetilde{\Upsilon}}$}$_{\text{FDM}}(\rho_{|\Phi _{\text{F}} \rangle}))$}. CRE approximations used $900$ decompositions.}
\label{fig:9}
\end{figure}
Here{\kern 0.9pt} in{\kern 0.9pt} \Fig{9}, we see the ratings of \smash{\scalebox{0.80}{$\hat{\widetilde{\Upsilon}}$}$_{\text{FDM}}$} are similar{\kern -0.75pt} to{\kern -0.75pt} those{\kern -0.75pt} of{\kern -0.75pt} \smash{\scalebox{0.80}{$\hat{\widetilde{\Upsilon}}$}$_{\text{FSM}}$} in \Fig{8}, except that here, the values for $\rho _{\text{GHZ} + 1}$ are much closer to the values for $\rho _{\text{F} + 1}$ (though still less than they are).  Thus, this test does not show any apparent problems, and exhibits the necessary features that neither $\rho _{2\text{-sep}}$ nor $\rho _{\text{MME} }$ can be considered free of $2$-partite entanglement correlations (see \App{H}).
\subsection{\label{sec:V.D}Comparison of all Mixed-Input Candidates}
The main difference between the GM measures (from \Eq{16} and \Eq{17}) and the SM and DM measures of \Eq{19} and \Eq{20} is that the GM measures tend to undervalue the amount of entanglement, which is a consequence of their interpretation of separability as the prime criterion for lack of entanglement.  The SM and DM measures take a more global approach, checking every possible $k$-partition for the presence of entanglement correlations, and as such, they correctly detect entanglement of every $k$-partitional type, in particular correctly not ignoring the maximal bipartite entanglement in \smash{$\rho_{\text{MME}}$}.

In the pure-input case, the FDM ent \smash{$\Upsilon_{\text{FDM}}$} was the only measure that could distinguish between \smash{$|\Phi _{\text{GHZ}} \rangle$} and \smash{$|\Phi _{\text{BP}} \rangle$} \textit{without} sacrificing detection of entanglement correlations, \textit{and} it equals the concurrence $C$ for two qubits.  Since $C$ for \textit{mixed} states is a convex-roof extension (CRE), and since \smash{$\hat{\Upsilon}_{\text{FDM}}$} is \textit{also} a CRE, then \smash{$\hat{\Upsilon}_{\text{FDM}}=C$} for \textit{mixed two-qubit states}, as well.
\section{\label{sec:VI}Ent-Concurrence}
\begin{figure}[H]
\centering
\vspace{-12pt}
\setlength{\unitlength}{0.01\linewidth}
\begin{picture}(100,0)
\put(1,24){\hypertarget{Sec:VI}{}}
\end{picture}
\end{figure}
\vspace{-39pt}
The ability of \smash{$\hat{\Upsilon}_{\text{FDM}}$} to detect and distinguish multipartite entanglement correlations and its link to $C$ both suggest that we adopt it as a universal measure of multipartite entanglement, called the \textit{ent-concurrence},
\begin{equation}
\hat{C}_{\Upsilon}(\rho)\equiv\hat{\Upsilon}_{\text{FDM}}(\rho),
\label{eq:22}
\end{equation}
where \smash{$\hat{\Upsilon}_{\text{FDM}}$} is defined in \Eq{20}, so that
\begin{equation}
\begin{array}{*{20}l}
   {\hat{C}_{\Upsilon}(\rho)} &\!\!{\kern -0.5pt} {\equiv\!\!{\kern 0.5pt} \mathop {\min }\limits_{\{ p_j ,\rho _j \} |_{\rho  = \sum\nolimits_j {p_j \rho _j } } } \!\!\!{\kern 0.5pt}\left(\!{\kern 0.5pt} {\sum\limits_j {\kern -0.5pt}{p_j\!{\kern -0.5pt}\sum\limits_{k = 2}^N \frac{1}{M_{k}}\sum\limits_{h = 1}^{\{_{\,k}^{N}\}}\!\! \sqrt{{\kern -1pt}\Upsilon^{(\mathbf{N}_h^{(\mathbf{k})} )}(\rho_{j} ){\kern -1pt}} } }{\kern 1pt} \right)\!{\kern -1pt},}  \\
\end{array}\!
\label{eq:23}
\end{equation}
where \smash{$M_{k}\equiv M_{\text{FDM}}M_{\text{DM}_{k}}$} is a normalization factor based on those of \Eq{10} and \Eq{11}, and \smash{$\{_{{\kern 1pt}k}^{N}\}$} are Stirling numbers of the second kind as in \Eq{9}, which is the number of different $N$-mode $k$-partitional ents \smash{$\Upsilon^{\shiftmath{-0.8pt}{(\mathbf{N}_{{\kern -0.5pt}\shiftmath{-0.5pt}{h}}^{{\kern 0.5pt}\shiftmath{-0.3pt}{(\mathbf{k})}} )}}(\rho )$}.

While $\hat{C}_{\Upsilon}$ detects all entanglement and distinguishes between different types of $\text{F}_k$-entanglement, it may also be useful to have a partition-specific view of how much entanglement exists between particular mode groups.  Therefore, in the notation of \cite{HedE}, we also define the \textit{$N$-mode partitional ent-concurrence vector} as
\begin{equation}
\Xi_{C_{\Upsilon}}^{(\mathbf{N})}(\rho)  \equiv\! \left( {\begin{array}{*{20}c}
   {\{ \hat{C}_{\Upsilon}^{(\mathbf{N}_h^{(\mathbf{2})} )}(\rho) \} }  \\
    \vdots   \\
   {\{ \hat{C}_{\Upsilon}^{(\mathbf{N}_h^{(\mathbf{N})} )}(\rho) \} }  \\
\end{array}} \right)\!,
\label{eq:24}
\end{equation}
where \smash{$\{ \hat{C}_{\Upsilon}^{(\mathbf{N}_h^{(\mathbf{k})} )}(\rho ) \}$} is the set of all $N$-mode $k$-partitional ent-concurrences of formation, the pure-input versions of which are in \Eq{13}. For example, in a $4$-mode system,
\begin{equation}
\scalebox{0.90}{$\begin{array}{l}
 \Xi_{C_\Upsilon  }^{(\mathbf{4})}  \equiv \Xi_{C_\Upsilon  }^{(1,2,3,4)}=  \\ 
 \rule{0pt}{28.5pt}\!\!\left(\!\!{\kern -0.6pt} {\begin{array}{*{20}c}
   {\hat{C}_\Upsilon ^{(1|2,\!3,\!4)} {\kern 1.0pt}\hat{C}_\Upsilon ^{(2|1,\!3,\!4)} {\kern 1.0pt}\hat{C}_\Upsilon ^{(3|1,\!2,\!4)} {\kern 1.0pt}\hat{C}_\Upsilon ^{(4|1,\!2,\!3)} {\kern 1.0pt}\hat{C}_\Upsilon ^{(1,\!2|3,\!4)} {\kern 1.0pt}\hat{C}_\Upsilon ^{(1,\!3|2,\!4)} {\kern 1.0pt}\hat{C}_\Upsilon ^{(1,\!4|2,\!3)} }  \\
   {\rule{0pt}{12.5pt}\!\hat{C}_\Upsilon ^{(1|2|3,\!4)} {\kern 1.0pt}\hat{C}_\Upsilon ^{(1|3|2,\!4)} {\kern 1.0pt}\hat{C}_\Upsilon ^{(1|4|2,\!3)} {\kern 1.0pt}\hat{C}_\Upsilon ^{(2|3|1,\!4)} {\kern 1.0pt}\hat{C}_\Upsilon ^{(2|4|1,\!3)} {\kern 1.0pt}\hat{C}_\Upsilon ^{(3|4|1,\!2)} }  \\
   {\rule{0pt}{12.5pt}\!\hat{C}_\Upsilon ^{(1|2|3|4)} }  \\
\end{array}}\!\! \right)\!{\kern -1pt}, \\ 
 \end{array}$}\!\!
\label{eq:25}
\end{equation}
where, for instance, 
\begin{equation}
\hat{C}_\Upsilon ^{(2|1,3,4)}  = \!\mathop {\min }\limits_{\{ p_j ,\rho _j \} |_{\rho  = \sum\nolimits_j {p_j \rho _j } } } \!\!\left( {\sum\nolimits_j {p_j \sqrt {\Upsilon ^{(2|1,3,4)} (\rho _j )} } }\, \right)\!.
\label{eq:26}
\end{equation}
Thus, the top row of \smash{$\Xi_{\shiftmath{0.5pt}{{\kern 0.5pt}C_{\Upsilon}}}^{\shiftmath{2.5pt}{(\mathbf{N})}}$} lists all contributions to $\text{F}_2$-entanglement, and so-on until the lowest row gives the $\text{F}_N$-entanglement, yielding a fine-grained view of the entanglement between each possible mode group.

For an intermediate view of entanglement, we can define the \textit{$N$-mode $k$-ent-concurrences of formation} as
\begin{equation}
\hat{C}_{\Upsilon_{k}}(\rho)\equiv \hat{\Upsilon}_{\text{DM}_{k}}(\rho),
\label{eq:27}
\end{equation}
which\hsp{-0.5} is\hsp{-0.5} a\hsp{-0.5} CRE\hsp{-0.5} of\hsp{-0.5} a\hsp{-0.5} $1$-norm\hsp{-0.5} over\hsp{-0.5} all\hsp{-0.5} \smash{$C_{\Upsilon}^{(\mathbf{N}_h^{(\mathbf{k})} )}(\rho )$}\hsp{-0.5} for\hsp{-0.5} a\hsp{-0.5} given\hsp{-0.5} $k$,\hsp{-0.5} as\hsp{-0.5} seen\hsp{-0.5} from\hsp{-0.5} \Eq{13}\hsp{-0.5} and\hsp{-0.5} the\hsp{-0.5} definition\hsp{-0.5} of\hsp{-0.5} \smash{$\Upsilon_{\text{DM}_{k}}(\rho)$}\hsp{-0.5} in\hsp{-0.5} \Eq{11}.
\subsection*{\label{sec:VI.A}Hierarchy of Maximally $\text{F}_N$-Entangled TGX States}
The ent-concurrence identifies a hierarchy among the maximally $\text{F}_N$-entangled states, which is easiest to see by examining its values for the subset of $\text{F}_N$-entangled TGX states from \App{G}, as in \Table{1}.
\begin{table}[H]
\caption{\label{tab:1}Normalized ent-concurrence $C_{\Upsilon}$ and normalized $k$-ent-concurrences $C_{\Upsilon_{k}}$ for each of the maximally $\text{F}_N$-entangled $4$-qubit{\kern -1pt} TGX{\kern -1pt} states{\kern -1pt} \smash{$|\Phi_{{j}}^{[L_{*}]}\rangle$} involving the first computational basis state \smash{$|1\rangle\equiv|1,\!1,\!1,\!1\rangle$}, from \App{G}, where $L_{*}$ is the number of levels with nonzero state coefficients. Normalizations are over these states alone, and may not be the normalizations over all states. The test states of \Eq{14} are \smash{$|\Phi_{\text{GHZ}}\rangle\equiv|\Phi_{1}^{[2]}\rangle$}, \smash{$|\Phi_{\text{BP}}\rangle\equiv|\Phi_{1}^{[4]}\rangle$}, and \smash{$|\Phi_{\text{F}}\rangle\equiv|\Phi_{2}^{[4]}\rangle$}, in the first three rows.}
\begin{ruledtabular}
\begin{tabular}{|c|c|c|c|c|}
$|\Phi_{j}^{[L_{*}]}\rangle$\rule{0pt}{10.5pt} & $C_{\Upsilon}$\hsp{13} & $C_{\Upsilon_{2}}$\hsp{13} & $C_{\Upsilon_{3}}$\hsp{13} & $C_{\Upsilon_{N}}$\hsp{13}\\[2.5pt]
\hline 
$|\Phi_{1}^{[2]}\rangle$\rule{0pt}{10.5pt} & $0.957$\hsp{13} & $0.946$\hsp{13} & $0.961$\hsp{13} & $1.000$\hsp{13} \\[2pt]
\hline 
$|\Phi_{1}^{[4]}\rangle$\rule{0pt}{10.5pt} & $0.922$\hsp{13} & $0.880$\hsp{13} & $0.957$\hsp{13} & $1.000$\hsp{13} \\[2pt]
\hline 
$|\Phi_{2}^{[4]}\rangle$\rule{0pt}{10.5pt} & $1.000$\hsp{13} & $1.000$\hsp{13} & $1.000$\hsp{13} & $1.000$\hsp{13} \\[2pt]
\hline 
$|\Phi_{3}^{[4]}\rangle$\rule{0pt}{10.5pt} & $0.922$\hsp{13} & $0.880$\hsp{13} & $0.957$\hsp{13} & $1.000$\hsp{13} \\[2pt]
\hline 
$|\Phi_{4}^{[4]}\rangle$\rule{0pt}{10.5pt} & $1.000$\hsp{13} & $1.000$\hsp{13} & $1.000$\hsp{13} & $1.000$\hsp{13} \\[2pt]
\hline 
$|\Phi_{5}^{[4]}\rangle$\rule{0pt}{10.5pt} & $0.922$\hsp{13} & $0.880$\hsp{13} & $0.957$\hsp{13} & $1.000$\hsp{13} \\[2pt]
\hline 
$|\Phi_{6}^{[4]}\rangle$\rule{0pt}{10.5pt} & $1.000$\hsp{13} & $1.000$\hsp{13} & $1.000$\hsp{13} & $1.000$\hsp{13} \\[2pt]
\hline 
$|\Phi_{7}^{[4]}\rangle$\rule{0pt}{10.5pt} & $1.000$\hsp{13} & $1.000$\hsp{13} & $1.000$\hsp{13} & $1.000$\hsp{13} \\[2pt]
\hline 
$|\Phi_{8}^{[4]}\rangle$\rule{0pt}{10.5pt} & $1.000$\hsp{13} & $1.000$\hsp{13} & $1.000$\hsp{13} & $1.000$\hsp{13} \\[2pt]
\hline 
$|\Phi_{9}^{[4]}\rangle$\rule{0pt}{10.5pt} & $1.000$\hsp{13} & $1.000$\hsp{13} & $1.000$\hsp{13} & $1.000$\hsp{13} \\[2pt]
\hline 
$|\Phi_{1}^{[6]}\rangle$\rule{0pt}{10.5pt} & $0.957$\hsp{13} & $0.946$\hsp{13} & $0.961$\hsp{13} & $1.000$\hsp{13} \\[2pt]
\hline 
$|\Phi_{2}^{[6]}\rangle$\rule{0pt}{10.5pt} & $0.957$\hsp{13} & $0.946$\hsp{13} & $0.961$\hsp{13} & $1.000$\hsp{13} \\[2pt]
\hline 
$|\Phi_{3}^{[6]}\rangle$\rule{0pt}{10.5pt} & $0.957$\hsp{13} & $0.946$\hsp{13} & $0.961$\hsp{13} & $1.000$\hsp{13} \\[2pt]
\hline 
$|\Phi_{1}^{[8]}\rangle$\rule{0pt}{10.5pt} & $0.957$\hsp{13} & $0.946$\hsp{13} & $0.961$\hsp{13} & $1.000$\hsp{13} \\
\end{tabular}
\end{ruledtabular}
\end{table}
In \Table{1}, the highest-rated maximally $\text{F}_N$-entangled TGX states, having \smash{$(C_{\Upsilon},\!C_{\Upsilon_{2}},\!C_{\Upsilon_{3}},\!C_{\Upsilon_{N}})=(1,\!1,\!1,\!1)$}, are the \textit{tier-1 $\text{F}_N$-entangled states},
\begin{equation}
\begin{array}{*{20}l}
   {|\Phi _2^{[4]} \rangle } &\!\!\! { =\! {\textstyle{1 \over {\sqrt 4 }}}(|1111\rangle  \!+\! |1122\rangle  \!+\! |2212\rangle  \!+\! |2221\rangle )}  \\
   {|\Phi _4^{[4]} \rangle } &\!\!\! { =\! {\textstyle{1 \over {\sqrt 4 }}}(|1111\rangle  \!+\! |1212\rangle  \!+\! |2122\rangle  \!+\! |2221\rangle )}  \\
   {|\Phi _6^{[4]} \rangle } &\!\!\! { =\! {\textstyle{1 \over {\sqrt 4 }}}(|1111\rangle  \!+\! |1221\rangle  \!+\! |2122\rangle  \!+\! |2212\rangle )}  \\
   {|\Phi _7^{[4]} \rangle } &\!\!\! { =\! {\textstyle{1 \over {\sqrt 4 }}}(|1111\rangle  \!+\! |1222\rangle  \!+\! |2112\rangle  \!+\! |2221\rangle )}  \\
   {|\Phi _8^{[4]} \rangle } &\!\!\! { =\! {\textstyle{1 \over {\sqrt 4 }}}(|1111\rangle  \!+\! |1222\rangle  \!+\! |2121\rangle  \!+\! |2212\rangle )}  \\
   {|\Phi _9^{[4]} \rangle } &\!\!\! { =\! {\textstyle{1 \over {\sqrt 4 }}}(|1111\rangle  \!+\! |1222\rangle  \!+\! |2122\rangle  \!+\! |2211\rangle ),}  \\
\end{array}
\label{eq:28}
\end{equation}
where $|abcd\rangle\equiv|a,\!b,\!c,\!d\rangle$.  The \textit{tier-2 $\text{F}_N$-entangled states}, with $(C_{\Upsilon},\!C_{\Upsilon_{2}},\!C_{\Upsilon_{3}},\!C_{\Upsilon_{N}})\approx(0.957,0.946,0.961,1)$, are
\begin{equation}
\scalebox{0.93}{$\begin{array}{*{20}l}
   {|\Phi _1^{[2]} \rangle } &\!\!\! { =\! {\textstyle{1 \over {\sqrt 2 }}}(|1111\rangle \! +\! |2222\rangle )}  \\
   {|\Phi _1^{[6]} \rangle } &\!\!\! { =\! {\textstyle{1 \over {\sqrt 6 }}}(|1111\rangle \! + \!|1122\rangle  \!+\! |1212\rangle \! + \!|2121\rangle  \!+\! |2211\rangle  \!+\! |2222\rangle )}  \\
   {|\Phi _2^{[6]} \rangle } &\!\!\! { =\! {\textstyle{1 \over {\sqrt 6 }}}(|1111\rangle  \!+\! |1122\rangle  \!+\! |1221\rangle  \!+\! |2112\rangle  \!+\! |2211\rangle  \!+\! |2222\rangle )}  \\
   {|\Phi _3^{[6]} \rangle } &\!\!\! { =\! {\textstyle{1 \over {\sqrt 6 }}}(|1111\rangle  \!+\! |1212\rangle  \!+\! |1221\rangle  \!+\! |2112\rangle  \!+\! |2121\rangle  \!+\! |2222\rangle )}  \\
   {|\Phi _1^{[8]} \rangle } &\!\!\! { =\! {\textstyle{1 \over {\sqrt 8 }}}(|1111\rangle  \!+\! |1122\rangle  \!+\! |1212\rangle  \!+\! |1221\rangle }\\  
{} &\!\!\! {{\kern 17pt}\!+\! |2112\rangle  \!+\! |2121\rangle  \!+\! |2211\rangle  \!+\! |2222\rangle ),}  \\
\end{array}$}
\label{eq:29}
\end{equation}
and\hsp{-0.7} the\hsp{-0.7} lowest-rated\hsp{-0.7} group,\hsp{-0.7} the\hsp{-0.7} \textit{tier-3\hsp{-0.7} $\text{F}_N$-entangled\hsp{-0.7} states},\hsp{-0.7} with $(C_{\Upsilon},\!C_{\Upsilon_{2}},\!C_{\Upsilon_{3}},\!C_{\Upsilon_{N}})\approx(0.922,0.880,0.957,1)$, are
\begin{equation}
\begin{array}{*{20}l}
   {|\Phi _1^{[4]} \rangle } &\!\!\! { =\! {\textstyle{1 \over {\sqrt 4 }}}(|1111\rangle  \!+\! |1122\rangle  \!+\! |2211\rangle  \!+\! |2222\rangle )}  \\
   {|\Phi _3^{[4]} \rangle } &\!\!\! { =\! {\textstyle{1 \over {\sqrt 4 }}}(|1111\rangle  \!+\! |1212\rangle  \!+\! |2121\rangle  \!+\! |2222\rangle )}  \\
   {|\Phi _5^{[4]} \rangle } &\!\!\! { =\! {\textstyle{1 \over {\sqrt 4 }}}(|1111\rangle  \!+\! |1221\rangle  \!+\! |2112\rangle  \!+\! |2222\rangle ).}  \\
\end{array}
\label{eq:30}
\end{equation}
All of these states are maximally $\text{F}_N$-entangled, as seen in \Table{1}, and furthermore, these are only a small portion of the ``phaseless'' maximally $\text{F}_N$-entangled TGX states, since similar sets can be generated by specifying a different common ``starting level'' than $|1\rangle=|1111\rangle$.

Whether or not other states exist that have higher ent-concurrence than the tier-1 states is still unknown, but none were found in the present numerical tests.

To see the most fine-grained view, from \Eq{25} the $N$-mode ent-concurrence vectors of \smash{$|\Phi _{\text{F}} \rangle\equiv|\Phi_{{\kern -0.5pt}2}^{{\kern 0.5pt}\shiftmath{0.5pt}{\scalemath{0.9}{[4]}}}\rangle$} (tier 1), \smash{$|\Phi_{\text{GHZ}}\rangle\equiv|\Phi_{{\kern -1pt}\shiftmath{-0.5pt}{1}}^{{\kern 0.5pt}\shiftmath{-0.0pt}{\scalemath{0.9}{[2]}}}\rangle$} (tier 2), and \smash{$|\Phi_{\text{BP}}\rangle\equiv|\Phi_{{\kern -1pt}\shiftmath{-0.5pt}{1}}^{{\kern 0.5pt}\shiftmath{-0.0pt}{\scalemath{0.9}{[4]}}}\rangle$} (tier 3) are
\begin{equation}
\Xi_{C_\Upsilon  }^{(\mathbf{N})} (\rho _{|\Phi _{\text{F}} \rangle } ) =\! \left(\!\! {\begin{array}{*{20}c}
   {\begin{array}{*{20}c}
   1 & 1 & 1 & 1 & {\sqrt {{\textstyle{2 \over 3}}} } & 1 & 1  \\
\end{array}}  \\
   {\begin{array}{*{20}c}
   {\sqrt {{\textstyle{8 \over 9}}} } & 1 & 1 & 1 & 1 & {\sqrt {{\textstyle{8 \over 9}}} }  \\
\end{array}}  \\
   1  \\
\end{array}}\!\! \right)\!,
\label{eq:31}
\end{equation}
\begin{equation}
\Xi_{C_\Upsilon  }^{(\mathbf{N})} (\rho _{|\Phi _{\text{GHZ}} \rangle } ) =\! \left(\!\! {\begin{array}{*{20}c}
   {\begin{array}{*{20}c}
   1 & 1 & 1 & 1 & {\sqrt {{\textstyle{2 \over 3}}} } & {\sqrt {{\textstyle{2 \over 3}}} } & {\sqrt {{\textstyle{2 \over 3}}} }  \\
\end{array}}  \\
   {\rule{0pt}{13pt}{\kern 0.1pt}\begin{array}{*{20}c}
   {\sqrt {{\textstyle{8 \over 9}}} } & {\sqrt {{\textstyle{8 \over 9}}} } & {\sqrt {{\textstyle{8 \over 9}}} } & {\sqrt {{\textstyle{8 \over 9}}} } & {\sqrt {{\textstyle{8 \over 9}}} } & {\sqrt {{\textstyle{8 \over 9}}} }  \\
\end{array}}  \\
   1  \\
\end{array}}\!\! \right)\!,
\label{eq:32}
\end{equation}
\begin{equation}
\Xi_{C_\Upsilon  }^{(\mathbf{N})} (\rho _{|\Phi _{\text{BP}} \rangle } ) =\! \left(\!\! {\begin{array}{*{20}c}
   {\begin{array}{*{20}c}
   1 & 1 & 1 & 1 & 0 & 1 & 1  \\
\end{array}}  \\
   {\begin{array}{*{20}c}
   {\sqrt {{\textstyle{2 \over 3}}} } & 1 & 1 & 1 & 1 & {\sqrt {{\textstyle{2 \over 3}}} }  \\
\end{array}}  \\
   1  \\
\end{array}}\!\! \right)\!.
\label{eq:33}
\end{equation}
The sum of all elements in each of \Eqs{31}{33} is the unnormalized{\kern -0.5pt} ent-concurrence,{\kern -0.5pt} yielding{\kern -0.5pt} $13.70$,{\kern -0.5pt} $13.11$,{\kern -0.5pt} and $12.63$, respectively,{\kern 0.5pt} which{\kern 0.5pt} are{\kern 0.5pt} the{\kern 0.5pt} first{\kern 0.5pt} three \smash{$\widetilde{\Upsilon}_{\text{FDM}}$} values (in a different order) in \Fig{5}.  In contrast, the square of all the elements in \Eqs{31}{33} (since these are pure states), yields the $N$-mode ent vectors of \cite{HedE}, the sums of which yield $13.44$, $12.33$, and $12.33$, respectively, which explains why the values of \smash{$\widetilde{\Upsilon}_{\text{FSM}}$} for \smash{$|\Phi_{\text{GHZ}}\rangle$} and \smash{$|\Phi_{\text{BP}}\rangle$} are equal in \Fig{4}, showing that the FSM measures were not able to able to distinguish these two states.

The worth of \smash{$\Xi_{\scalemath{0.9}{C_\Upsilon } }^{\scalemath{0.9}{(\mathbf{N})}}$} is that it shows us \textit{between which mode groups} entanglement and separability occur. For example, the $0$ in \Eq{33} shows that the mode groups defined by the partitioning $(1,2|3,4)$ are separable in \smash{$\rho _{\shiftmath{0.6pt}{|\Phi _{\text{BP}} \rangle} }$}, which is true since that is the Bell product, but \Eq{33} \textit{also} shows that the \textit{entanglement is maximal} between all other bipartitions of the state (seen in its top row). Thus, the ent-concurrence does not ignore all of these bipartite entanglement correlations just because one of them is zero, as the GM measures do.
\section{\label{sec:VII}Absolute Ent-Concurrence}
\begin{figure}[H]
\centering
\vspace{-12pt}
\setlength{\unitlength}{0.01\linewidth}
\begin{picture}(100,0)
\put(1,24){\hypertarget{Sec:VII}{}}
\end{picture}
\end{figure}
\vspace{-39pt}
While the ent-concurrence (and its accompanying notions of $N$-mode partitional ent-concurrence vector and $k$-ent-concurrence) evaluates the multipartite entanglement resources of \textit{the entire input state in its full space}, another dimension of details can be gleaned by evaluating the entanglement \textit{within reductions of the input state}.

Thus for mode group $\mathbf{m}$, the \textit{$S$-mode partitional ent-concurrence vector} is
\begin{equation}
\Xi_{C_{\Upsilon}} ^{(\mathbf{m})}(\rho)  \equiv\! \left( {\begin{array}{*{20}c}
   {\{ \hat{C}_{\Upsilon}^{(\mathbf{m}_h^{(\mathbf{2})} )}(\rho) \} }  \\
    \vdots   \\
   {\{ \hat{C}_{\Upsilon}^{(\mathbf{m}_h^{(\mathbf{S})} )}(\rho) \} }  \\
\end{array}} \right)\!,
\label{eq:34}
\end{equation}
where $\mathbf{m}\equiv(m_{1},\ldots,m_{S});\, S\in 2,\ldots,N$ are the modes to which $\rho\equiv\rho^{(1,\ldots,N)}$ is being reduced before being partitioned, and \smash{$\{ \hat{C}_{{\kern -1pt}\shiftmath{-0.9pt}{{\kern 1.0pt}\scalemath{0.85}{\Upsilon}}}^{{\kern 0.0pt}\shiftmath{-0.1pt}{(\mathbf{m}_{{\kern -0.1pt}\shiftmath{-0.6pt}{h}}^{{\kern 0.5pt}\shiftmath{-0.3pt}{(\mathbf{T})}} )}}\}$} is the set of all $S$-mode $T$-partitional ent-concurrences of a given reduction \smash{$\redx{\rho}{\mathbf{m}}$} where $T\in 2,\ldots,S$, where for a particular partitioning labeled by $h$, the \textit{partitional ent-concurrence} is
\begin{equation}
\hat{C}_\Upsilon ^{(\mathbf{m}_{h}^{(\mathbf{T})})}(\rho) \!{\kern 0.5pt} \equiv \!\mathop {\min }\limits_{\{ p_j ,\rho _j \} |_{\rho  = \sum\nolimits_j \!{p_j \rho _j } } } \!\!\left( {\sum\nolimits_j {p_j \sqrt {\Upsilon ^{(\mathbf{m}_{h}^{(\mathbf{T})})} (\rho _j )} } }\, \right)\!,
\label{eq:35}
\end{equation}
where \smash{$\Upsilon ^{(\mathbf{m}_{h}^{(\mathbf{T})})}$} is the \textit{partitional ent} (of $h$-labeled partition \smash{$\mathbf{m}_{h}^{{\kern 1.5pt}\shiftmath{0pt}{(\mathbf{T})}}$}) mentioned after \Eq{5} and defined in detail in \cite{HedE}.  Thus,{\kern -1pt} row{\kern -1pt} $T{\kern -1.5pt}-{\kern -1pt}1${\kern -1pt} of \smash{$\Xi_{C_{\Upsilon}} ^{\shiftmath{0.6pt}{(\mathbf{m})}}$}{\kern -1pt} lists{\kern -1pt} all{\kern -1pt} possible{\kern -1pt} $T$-partitional ent-concurrences of the mode-$\mathbf{m}$ reduction of $\rho$.

Since a partitional ent-concurrence vector exists for each reduction $\mathbf{m}$, we can define the \textit{ent-concurrence array} as the matrix \smash{$\widetilde{\nabla}_{C_{\Upsilon}}$} (not a gradient) whose elements are partitional ent-concurrence vectors,
\begin{equation}
{(\widetilde{\nabla}_{C_{\Upsilon}})}_{k,l} (\rho ) \equiv \Xi_{C_{\Upsilon}} ^{((\text{nCk}(\mathbf{c},k))_{l, \cdots } )},
\label{eq:36}
\end{equation}
where \smash{$\mathbf{c} \equiv (1, \ldots ,N)$}, \smash{$k\in 2,\ldots,N$}, and \smash{$l \in 1, \ldots ,\binom{N}{k}$} where \smash{$\binom{N}{k}{\kern 0.5pt}\equiv{\kern 0.5pt}\shiftmath{0.5pt}{\scalemath{0.92}{\frac{N!}{k!(N-k)!}}}$}, and \smash{$\text{nCk}(\mathbf{v},k)$} is the vectorized $n$-choose-$k$ function yielding the matrix whose rows are each unique combinations of the elements of $\mathbf{v}$ chosen $k$ at a time, and \smash{$A_{l,\cdots}$} is the $l$th row of matrix $A$.  For example in $N=4$, (suppressing input arguments $\rho$)
\begin{equation}
\widetilde{\nabla}_{C_{\Upsilon}} (\rho ) =\! \left(\!\! {\begin{array}{*{20}c}
   {\begin{array}{*{20}c}
   {\Xi_{C_{\Upsilon}} ^{(1,2)} } & {\Xi_{C_{\Upsilon}} ^{(1,3)} } & {\Xi_{C_{\Upsilon}} ^{(1,4)} } & {\Xi_{C_{\Upsilon}} ^{(2,3)} } & {\Xi_{C_{\Upsilon}} ^{(2,4)} } & {\Xi_{C_{\Upsilon}} ^{(3,4)} }  \\
\end{array}}  \\
   {\begin{array}{*{20}c}
   {\rule{0pt}{12pt}\!\Xi_{C_{\Upsilon}} ^{(1,2,3)} } & {\Xi_{C_{\Upsilon}} ^{(1,2,4)} } & {\Xi_{C_{\Upsilon}} ^{(1,3,4)} } & {\Xi_{C_{\Upsilon}} ^{(2,3,4)} }  \\
\end{array}}  \\
   {\rule{0pt}{12pt}\!\Xi_{C_{\Upsilon}} ^{(1,2,3,4)} }  \\
\end{array}}\!\!\! \right)\!,
\label{eq:37}
\end{equation}
where the $2$-mode partitional ent-concurrence vectors have just one element, such as
\begin{equation}
\Xi_{C_{\Upsilon}} ^{(2,3)} =\hat{C}_\Upsilon ^{(2|3)}= \hat{C}_\Upsilon ^{(2,3)},
\label{eq:38}
\end{equation}
and $3$-mode partitional ent-concurrence vectors look like
\begin{equation}
\Xi_{C_{\Upsilon}} ^{(1,2,4)} \! =\! \left(\!\! {\begin{array}{*{20}c}
   {\begin{array}{*{20}c}
   {\hat{C}_\Upsilon ^{(1|2,4)} } & {\hat{C}_\Upsilon^{(2|1,4)} } & {\hat{C}_\Upsilon^{(4|1,2)} }  \\
\end{array}}  \\
   {\hat{C}_\Upsilon ^{(1|2|4)} }  \\
\end{array}}\!\! \right)\!,
\label{eq:39}
\end{equation}
and the $4$-mode partitional ent-concurrence vector \smash{$\Xi_{C_{\Upsilon}} ^{(1,2,3,4)}$} is given by \Eq{25}.

Then, to create a universal multipartite entanglement measure that can detect \textit{all possible} entanglement correlations of a state \textit{including those of all of its reductions}, we can define the \textit{absolute ent-concurrence} as
\begin{equation}
C_{\Upsilon_{\text{abs}}} (\rho ) \equiv \frac{{||\widetilde{\nabla}_{C_{\Upsilon}} (\rho )||_{1} }}{{\max (||\widetilde{\nabla}_{C_{\Upsilon}}(\rho ) ||_{1})}}\,,
\label{eq:40}
\end{equation}
which is the $1$-norm over all partitional ent-concurrences normalized to its maximum value over all input states.

Thus, by \hyperlink{Thm:1}{Theorem 1} from \App{D}, \smash{$C_{\Upsilon_{\text{abs}}}$} captures a measurement of all possible ways in which a state can be entanglement-correlated.\hsp{-0.2} 
The\hsp{-0.2} main\hsp{-0.2} drawback\hsp{-0.2} of\hsp{-0.2} the\hsp{-0.2} absolute\hsp{-0.2} ent-concurrence\hsp{-0.2} of\hsp{-0.2} \Eq{40}\hsp{-0.2} is\hsp{-0.2} that\hsp{-0.2} it\hsp{-0.2} generally\hsp{-0.2} requires convex-roof extensions (CREs) in all elements of \smash{$\widetilde{\nabla}_{{\kern -1.5pt}C_{\Upsilon}}$}{\kern -0.5pt}, \textit{even when the input is pure}, since \textit{reductions} of the pure decomposition states $\rho_j$ of $\rho$ are generally mixed.  This means that it is usually computationally intractable to calculate \smash{$C_{\Upsilon_{\text{abs}}}$}, even for pure $\rho$.

For states like $|\psi_{\text{W}}\rangle$, where one of its prime characteristics is that it retains entanglement after the removal of a particle (tracing away a mode) \cite{DuVC}, finding the entanglement of its \textit{reductions} is possible with the absolute ent-concurrence, and the ent-concurrence array of \Eq{36} is an excellent tool for a high-resolution picture of all possible entanglement correlations of the state.  These measures would certainly show exactly how $|\psi_{\text{W}}\rangle$ differs from $|\Phi_{\text{GHZ}}\rangle$, which is separable after removal of any particles. For example, the ent-concurrence array of $|\psi_{\text{W}}\rangle$ is
\begin{equation}
\begin{array}{l}
 \widetilde{\nabla}_{C_{\Upsilon}} (\rho_{|\psi_{\text{W}}\rangle} )=  \\ 
 {\left(\!\!\!\!\! {\begin{array}{*{20}c}
   {\begin{array}{*{20}c}
   {({\textstyle{1 \over 2}})} & {({\textstyle{1 \over 2}})} & {({\textstyle{1 \over 2}})} & {({\textstyle{1 \over 2}})} & {({\textstyle{1 \over 2}})} & {({\textstyle{1 \over 2}})}  \\
\end{array}}  \\
   {\rule{0pt}{23pt}{\kern -0.1pt}\begin{array}{*{20}c}
   {\left(\!\!\!\! {\begin{array}{*{20}c}
   {\sqrt{\!\frac{1}{2}\!}\,\sqrt{\!\frac{1}{2}\!}\,\sqrt{\!\frac{1}{2}\!}}  \\
   \sqrt{\!\frac{1}{2}\!}  \\
\end{array}}\! \right)} &\!\!\!\!\! {\left(\!\!\!\! {\begin{array}{*{20}c}
   {\sqrt{\!\frac{1}{2}\!}\,\sqrt{\!\frac{1}{2}\!}\,\sqrt{\!\frac{1}{2}\!}}  \\
   \sqrt{\!\frac{1}{2}\!}  \\
\end{array}}\! \right)} &\!\!\!\!\! {\left(\!\!\!\! {\begin{array}{*{20}c}
   {\sqrt{\!\frac{1}{2}\!}\,\sqrt{\!\frac{1}{2}\!}\,\sqrt{\!\frac{1}{2}\!}}  \\
   \sqrt{\!\frac{1}{2}\!}  \\
\end{array}}\! \right)} &\!\!\!\!\! {\left(\!\!\!\! {\begin{array}{*{20}c}
   {\sqrt{\!\frac{1}{2}\!}\,\sqrt{\!\frac{1}{2}\!}\,\sqrt{\!\frac{1}{2}\!}}  \\
   \sqrt{\!\frac{1}{2}\!}  \\
\end{array}}\! \right)}  \\
\end{array}}  \\
   {\rule{0pt}{33pt}{\kern -0.1pt}\left(\!\!\!\! {\begin{array}{*{20}c}
   {\begin{array}{*{20}c}
   \sqrt{\!\frac{3}{4}\!} &\!\! \sqrt{\!\frac{3}{4}\!} &\!\! \sqrt{\!\frac{3}{4}\!} &\!\! \sqrt{\!\frac{3}{4}\!} &\!\! \sqrt{\!\frac{2}{3}\!} &\!\! \sqrt{\!\frac{2}{3}\!} &\!\! \sqrt{\!\frac{2}{3}\!}  \\
\end{array}}  \\
   {\rule{0pt}{13pt}{\kern -0.1pt}\begin{array}{*{20}c}
   \sqrt{\!\frac{13}{18}\!} &\!\! \sqrt{\!\frac{13}{18}\!} &\!\! \sqrt{\!\frac{13}{18}\!} &\!\! \sqrt{\!\frac{13}{18}\!} &\!\! \sqrt{\!\frac{13}{18}\!} &\!\! \sqrt{\!\frac{13}{18}\!}  \\
\end{array}}  \\
   \rule{0pt}{14pt}{\kern -0.1pt}\sqrt{\!\frac{3}{4}\!}  \\
\end{array}}\!\!\!\! \right)}  \\
\end{array}}\!\!\!\!\! \right)\!,} \\ 
\end{array}
\label{eq:41}
\end{equation}
while for $|\Phi_{\text{GHZ}}\rangle$, we have
\begin{equation}
\begin{array}{l}
 \widetilde{\nabla}_{C_{\Upsilon}} (\rho_{|\Phi_{\text{GHZ}}\rangle} )=  \\ 
 {\left(\!\!\!\!\! {\begin{array}{*{20}c}
   {\begin{array}{*{20}c}
   {(0)} & {(0)} & {(0)} & {(0)} & {(0)} & {(0)}  \\
\end{array}}  \\
   {\begin{array}{*{20}c}
   {\left( {\begin{array}{*{20}c}
   {0\,\,0\,\,0}  \\
   0  \\
\end{array}} \right)} &\!\!\! {\left( {\begin{array}{*{20}c}
   {0\,\,0\,\,0}  \\
   0  \\
\end{array}} \right)} &\!\!\! {\left( {\begin{array}{*{20}c}
   {0\,\,0\,\,0}  \\
   0  \\
\end{array}} \right)} &\!\!\! {\left( {\begin{array}{*{20}c}
   {0\,\,0\,\,0}  \\
   0  \\
\end{array}} \right)}  \\
\end{array}}  \\
   {\rule{0pt}{30pt}{\kern -0.1pt}\left(\!\!\!\! {\begin{array}{*{20}c}
   {\begin{array}{*{20}c}
   1 & 1 & 1 & 1 & {\sqrt {\!{\textstyle{2 \over 3}}\!} } & {\sqrt {\!{\textstyle{2 \over 3}}\!} } & {\sqrt {\!{\textstyle{2 \over 3}}\!} }  \\
\end{array}}  \\
   {\begin{array}{*{20}c}
   {\sqrt {\!{\textstyle{8 \over 9}}\!} } & {\sqrt {\!{\textstyle{8 \over 9}}\!} } & {\sqrt {\!{\textstyle{8 \over 9}}\!} } & {\sqrt {\!{\textstyle{8 \over 9}}\!} } & {\sqrt {\!{\textstyle{8 \over 9}}\!} } & {\sqrt {\!{\textstyle{8 \over 9}}\!} }  \\
\end{array}}  \\
   1  \\
\end{array}}\!\!\!\! \right)}  \\
\end{array}}\!\!\!\!\! \right)\!,} \\ 
\end{array}
\label{eq:42}
\end{equation}
showing that \smash{$|\psi_{\text{W}}\rangle$} has many more sites of entanglement correlation than \smash{$|\Phi_{\text{GHZ}}\rangle$}, as shown also by their unnormalized absolute ent-concurrences, \smash{$\widetilde{C}_{\Upsilon_{\text{abs}}} (\rho_{|\psi_{\text{W}}\rangle} ){\kern -1pt}\approx{\kern -1pt} 26.19$} and \smash{$\widetilde{C}_{\Upsilon_{\text{abs}}} (\rho_{\shiftmath{0.5pt}{|\Phi_{\text{GHZ}}\rangle}} ){\kern -1pt}\approx{\kern -1pt} 13.11$}, although \smash{$|\psi_{\text{W}}\rangle$} contains no mode groups that are maximally entangled (since none get up to $1$, since each element \textit{is} normalized), while \smash{$|\Phi_{\text{GHZ}}\rangle$} contains \textit{five} maximally entangled mode groups.

\textit{However}, as pointed out in \cite{HedE}, it is important to keep in mind that these entanglement correlations may not all be simultaneously available as resources.  Rather, the ent-concurrence array shows us all \textit{potential} entanglement resources a state has to offer.  Therefore, whether we consider \smash{$|\psi_{\text{W}}\rangle$} to be more or less ``entangled'' than \smash{$|\Phi_{\text{GHZ}}\rangle$} depends on our specific application, but the ent-concurrence array gives us a tool for assessing this.

For comparison, for \smash{$|\Phi_{\text{F}}\rangle$} of \Eq{14},
\begin{equation}
\begin{array}{l}
 \widetilde{\nabla}_{C_{\Upsilon}} (\rho_{|\Phi_{\text{F}}\rangle} )=  \\ 
 {\left(\!\!\!\!\! {\begin{array}{*{20}c}
   {\begin{array}{*{20}c}
   {(0)} & {(0)} & {(0)} & {(0)} & {(0)} & {(0.0541)}  \\
\end{array}}  \\
   {\begin{array}{*{20}c}
   {\left(\!\!\! {\begin{array}{*{20}c}
   {1\,\,1\,\,0.0541}  \\
   0.817  \\
\end{array}}\!\! \right)} &\!\!\!\! {\left(\!\!\! {\begin{array}{*{20}c}
   {1\,\,1\,\,0.0541}  \\
   0.817  \\
\end{array}}\!\! \right)} &\!\!\!\! {\left(\!\! {\begin{array}{*{20}c}
   {0\,\,1\,\,1}  \\
   \sqrt{\!\frac{2}{3}\!}  \\
\end{array}}\! \right)} &\!\!\!\! {\left(\!\! {\begin{array}{*{20}c}
   {0\,\,1\,\,1}  \\
   \sqrt{\!\frac{2}{3}\!}  \\
\end{array}}\! \right)}  \\
\end{array}}  \\
   {\left(\!\!\! {\begin{array}{*{20}c}
   {\begin{array}{*{20}c}
   1 &\!\! 1 &\!\! 1 &\!\! 1 &\!\! \sqrt{\!\frac{2}{3}\!} &\!\! 1 &\!\! 1  \\
\end{array}}  \\
   {\begin{array}{*{20}c}
   \sqrt{\!\frac{8}{9}\!} &\!\! 1 &\!\! 1 &\!\! 1 &\!\! 1 &\!\! \sqrt{\!\frac{8}{9}\!}  \\
\end{array}}  \\
   1  \\
\end{array}}\!\!\! \right)}  \\
\end{array}}\!\!\!\!\! \right)\!,} \\ 
\end{array}
\label{eq:43}
\end{equation}
so that \smash{$\widetilde{C}_{\Upsilon_{\text{abs}}} (\rho_{|\Phi_{\text{F}}\rangle} )\approx 25.13$}, and for \smash{$|\Phi_{\text{BP}}\rangle$} of \Eq{14},
\begin{equation}
\begin{array}{l}
 \widetilde{\nabla}_{C_{\Upsilon}} (\rho_{|\Phi_{\text{BP}}\rangle} )=  \\ 
 {\left(\!\!\!\!\! {\begin{array}{*{20}c}
   {\begin{array}{*{20}c}
   (1) & {(0)} & {(0)} & {(0)} & {(0)} & {(1)}  \\
\end{array}}  \\
   {\begin{array}{*{20}c}
   {\left(\!\! {\begin{array}{*{20}c}
   {1\,\,1\,\,0}  \\
   \sqrt{\!\frac{2}{3}\!}  \\
\end{array}}\! \right)} &\!\!\!\! {\left(\!\! {\begin{array}{*{20}c}
   {1\,\,1\,\,0}  \\
   \sqrt{\!\frac{2}{3}\!}  \\
\end{array}}\! \right)} &\!\!\!\! {\left(\!\! {\begin{array}{*{20}c}
   {0\,\,1\,\,1}  \\
   \sqrt{\!\frac{2}{3}\!}  \\
\end{array}}\! \right)} &\!\!\!\! {\left(\!\! {\begin{array}{*{20}c}
   {0\,\,1\,\,1}  \\
   \sqrt{\!\frac{2}{3}\!}  \\
\end{array}}\! \right)}  \\
\end{array}}  \\
   {\left(\!\!\!\! {\begin{array}{*{20}c}
   {\begin{array}{*{20}c}
   1 &\!\! 1 &\!\! 1 &\!\! 1 &\!\! 0 &\!\! 1 &\!\! 1  \\
\end{array}}  \\
   {\begin{array}{*{20}c}
   \sqrt{\!\frac{2}{3}\!} &\!\! 1 &\!\! 1 &\!\! 1 &\!\! 1 &\!\! \sqrt{\!\frac{2}{3}\!}  \\
\end{array}}  \\
   1  \\
\end{array}}\!\!\! \right)}  \\
\end{array}}\!\!\!\!\! \right)\!,} \\ 
\end{array}
\label{eq:44}
\end{equation}
so that \smash{$\widetilde{C}_{\Upsilon_{\text{abs}}} (\rho_{|\Phi_{\text{BP}}\rangle} )\!\approx\! 25.90$}.  Thus, $|\Phi_{\text{BP}}\rangle$ actually has the most occurrences of maximally-entangled mode groups with $21$ of them, while $|\Phi_{\text{F}}\rangle$ has the next highest number at $19$ of them.  These two states also contain reductions that are MME states as introduced in the text after \Eq{21}, and $|\Phi_{\text{F}}\rangle$ and $|\Phi_{\text{BP}}\rangle$ also have some rank-$4$ reductions that were luckily diagonal and therefore separable by any measure. Thus, \smash{$C_{\Upsilon_{\text{abs}}}$} orders states \textit{differently} than \smash{$C_{\Upsilon}$} due to its inclusion of the reductions.
\subsection*{\label{sec:VII.A}A Possible RMS Relationship}
From \Eqs{41}{44}, in each $S$-mode ent-concurrence vector \smash{$\Xi_{C_{\Upsilon}}^{(\mathbf{m})}$}, the higher-partitional elements are the root-mean-square (RMS) of 2-partitional elements with one partition{\kern -1pt} of{\kern -1pt} the{\kern -1pt} same{\kern -1pt} mode{\kern -1pt} group.{\kern -1pt} Thus,{\kern -1pt} for{\kern -1pt} the{\kern -1pt} $3$-mode{\kern -1pt} vectors,
\begin{equation}
\scalebox{0.97}{$\hat C_\Upsilon ^{(a|b|c)}  = \text{rms}(\hat C_\Upsilon ^{(a|b,c)} ,\hat C_\Upsilon ^{(b|a,c)} ,\hat C_\Upsilon ^{(c|a,b)} ),$}
\label{eq:45}
\end{equation}
where \smash{$\text{rms}(\mathbf{x}) \equiv (\shiftmath{1.8pt}{\scalemath{0.95}{\textstyle{1 \over {\dim (\mathbf{x})}}}}{\kern -0.5pt}\sum\nolimits_{\shiftmath{0.5pt}{k\! =\! 1}}^{\dim (\mathbf{x})} {{\kern -0.5pt}|x_k |^2 } )^{1/2}$}. For example, in \Eq{43}, \smash{$\hat C_{{\kern 0.5pt}\shiftmath{-1.5pt}{\scalemath{0.9}{\Upsilon}}} ^{\shiftmath{-0.5pt}{(1|2|3)}} = [\scalemath{0.9}{\textstyle{1 \over 3}}(1^{\shiftmath{-1.0pt}{2}}  + 1^{\shiftmath{-1.0pt}{2}}  + 0.0541^{\shiftmath{-1.0pt}{2}} )]^{\shiftmath{-1.0pt}{1/2}}= 0.817$}.  For the{\kern 1pt} $N$-mode{\kern 1pt} vectors,
\begin{equation}
\scalebox{0.97}{$\begin{array}{*{20}l}
   {\hat C_\Upsilon ^{(a|b|c,d)} } &\!\!\! { = \text{rms}(\hat C_\Upsilon ^{(a|b,c,d)} ,\hat C_\Upsilon ^{(b|a,c,d)} ,\hat C_\Upsilon ^{(a,b|c,d)} ),}  \\
   {\rule{0pt}{12pt}{\kern 0.1pt}\hat C_\Upsilon ^{(1|2|3|4)} } &\!\!\! { = \text{rms}(\hat C_\Upsilon ^{(1|2,3,4)} ,\hat C_\Upsilon ^{(2|1,3,4)} ,\hat C_\Upsilon ^{(3|1,2,4)} ,\hat C_\Upsilon ^{(4|1,2,3)} ),}  \\
\end{array}$}
\label{eq:46}
\end{equation}
up to irrelevant mode permutations in and of the mode groups. Since this was only tested for \Eqs{41}{44}, it is merely a hypothetical relationship in general at this time.
\section{\label{sec:VIII}Conclusions}
\begin{figure}[H]
\centering
\vspace{-12pt}
\setlength{\unitlength}{0.01\linewidth}
\begin{picture}(100,0)
\put(1,24){\hypertarget{Sec:VIII}{}}
\end{picture}
\end{figure}
\vspace{-39pt}
We have explored several multipartite entanglement measures, and found that one of them called the \textit{ent-concurrence} $\hat{C}_{\Upsilon}$ of \Eq{22}, can distinguish between maximally $\text{F}_N$-entangled states with different amounts of entanglement between all possible mode groups. Its name indicates the fact that $\hat{C}_{\Upsilon}$ is exactly equal to the concurrence $C$ for all pure and mixed states of two qubits, while for all larger systems, it is a function of \textit{the ent} $\Upsilon$, a necessary and sufficient measure of $\text{F}_N$-entanglement in $N$-mode systems \cite{HedE,HedD}. 

The reason that an $\text{F}_N$-entanglement measure like the ent $\Upsilon$ can be used in a  multipartite entanglement measure is that since every partitioning of $T\leq N$ mode groups can be treated as a $T$-mode system, $\Upsilon$ can be adapted as the \textit{$N$-mode partitional ent} \smash{${\Upsilon}^{\shiftmath{-1.0pt}{(\mathbf{N}^{\shiftmath{-1.5pt}{(\mathbf{T})}})}}$} to evaluate the $\text{F}_T$-entanglement of those $T$ mode groups. Then, the $N$-mode partitional ent-concurrence \smash{$ C_{{\kern -1pt}\shiftmath{-1.1pt}{{\kern 1.0pt}\scalemath{0.85}{\Upsilon}}}^{{\kern 0.0pt}\shiftmath{-0.3pt}{(\mathbf{N}^{{\kern 0.5pt}\shiftmath{-1.5pt}{(\mathbf{T})}} )}}$} for each partitioning of the state is the square root of the $N$-mode partitional ent, and the ent-concurrence $C_{\Upsilon}$ is a $1$-norm over all of the $N$-mode partitional ent-concurrences.  Mixed states are handled by the convex-roof extension (CRE) of $C_{\Upsilon}$ as $\hat{C}_{\Upsilon}$ (as with $C$, we can omit the hat).

Note\hsp{-0.5} that\hsp{-0.5} for\hsp{-0.5} pure\hsp{-0.5} and\hsp{-0.5} mixed\hsp{-0.5} states\hsp{-0.5} of\hsp{-0.5} a\hsp{-0.5} $2$-qubit\hsp{-0.5} system,\hsp{-0.5} \smash{$(\hat{C}_{\Upsilon})^2$}\hsp{-0.5} is\hsp{-0.5} exactly\hsp{-0.5} equal\hsp{-0.5} to\hsp{-0.5} the\hsp{-0.5} \textit{tangle}\hsp{-0.5} from\hsp{-0.5} \cite{CoKW}.\hsp{-0.5}  Thus, \smash{$\hat{C}_{\Upsilon}$} can be thought of as the square root of a multipartite generalization of the tangle.  Then, while adding up all the $N$-mode partitional ents yielded the occasional inability to distinguish between different kinds of $\text{F}_k$-entanglement for pure states, we found that the ent-concurrence $C_{\Upsilon}$ did not have this problem, since it is sensitive to how the entanglement is distributed throughout\hsp{-1} $\rho$.  Therefore, \smash{$\hat{C}_{\Upsilon}$} is a more appropriate measure than\hsp{1} \smash{$(\hat{C}_{\Upsilon})^2$} as a generalization of the tangle, despite their close relationship.

While ent-concurrence $C_{\Upsilon}$ is a good measure of multipartite entanglement for applications needing entanglement of the full input state $\rho$, we also defined the \textit{absolute ent-concurrence} $C_{\Upsilon_{\text{abs}}}$ as a measure that additionally takes into account the entanglement between all possible partitions of all possible \textit{reductions} $\redx{\rho}{\mathbf{m}}$. Thus, $C_{\Upsilon_{\text{abs}}}$ is for applications where it is important to have entanglement within the reductions as well as the full state.

The ent-concurrence array \smash{$\widetilde{\nabla}_{C_{\Upsilon}}$} of \Eq{36} shows that our application really determines what kind of measure we use. One alternative not explored here is to just target a specific reduction or a group of these such as all reductions composed of three modes, and define a \textit{modal ent-concurrence vector} and accompanying measure of \textit{modal ent-concurrence} in analogy to the modal ent of \cite{HedE}. Such measures could ignore entanglement in the full Hilbert space, focusing only on entanglement in the reductions.  Since this kind of measure is very specific, it allows one to make highly customized multipartite entanglement measures suited to specific applications.

We also noted that entanglement means \textit{nonseparable nonlocal correlation} (NSNLC), and that measures, such as ``genuinely multipartite'' (GM) entanglement measures, which test for the presence of \textit{any separability}, are not sufficient to detect the absence of all NSNLC, and are thus insufficient to detect all multipartite entanglement.

However, GM entanglement measures may still have use as \textit{$k$-inseparability measures}, allowing us to determine whether separability is possible between any $k$-mode groups.  Yet we must keep in mind that just because a state is separable over a \textit{particular} $k$-partition does not mean that NSNLC (and thus entanglement) do not exist between \textit{other} $k$-partitions.  For the purpose of $k$-inseparability measures, our proposed \textit{strict} FGM (SFGM) of formation from \Eq{17} gives the option to only report as $k$-separable those mixed states for which every pure decomposition state is separable over the same particular mode groups, to avoid the fallacy of thinking of states such as \Eq{2} as being devoid of $k$-partitional NSNLC, as explained in \App{H}.

Another important point, made in \App{B}, is that true separability implies the ability to mathematically reconstruct a mixed parent state from its reductions.  Therefore, if a measure reports any state as separable in some way, there must be, at least in principle, a way to find the decompositions of the relevant reduced states that can be used to exactly reconstruct the parent state. In general, GM measures do not indicate whether such reconstructions are possible, while the ent-concurrence (and a few other proposed measures) are \textit{guaranteed} to indicate this.

While the ent-concurrence $C_{\Upsilon}$ is a simple and easily computable multipartite entanglement measure for pure states, its mixed-state definition as a convex-roof extension (CRE) makes it intractable to approximate for states with rank $>2$, a difficulty common to all CRE-based measures.  Therefore, an interesting avenue for future research is the search for a \textit{computable} formula for the ent-concurrence $C_{\Upsilon}$ for all mixed-state input.

We may also conjecture that different ``tiers'' of maximally $\text{F}_N$-entangled states (such as those in \Sec{VI}) contain enough states to make a maximally-entangled basis (MEB) set.  This was proven to be always possible in \cite{HedE} for $\text{F}_N$-entangled states, so the tier-specific version can be considered as the ``strong-MEB'' theorem, and would also be interesting for future research.

Hopefully, the ent-concurrence $C_{\Upsilon}$ will enable advancements in the study of multipartite entanglement and our understanding of it.
\vspace{-4pt}
\begin{acknowledgments}
Many thanks to Ting Yu and B.{\kern 2.5pt}D.{\kern 2.5pt}Clader for helpful feedback and discussions.
\end{acknowledgments}
\begin{appendix}
\vspace{-4pt}
\section{\label{sec:App.A}Brief Review of Reduced States}
\begin{figure}[H]
\centering
\vspace{-12pt}
\setlength{\unitlength}{0.01\linewidth}
\begin{picture}(100,0)
\put(1,24){\hypertarget{Sec:IX}{}\hypertarget{Sec:App.A}{}}
\end{picture}
\end{figure}
\vspace{-39pt}
We represent \textit{multipartite state reduction} to a composite subsystem of $S\in 1,\ldots,N$ potentially noncontiguous and reordered modes $\mathbf{m} \equiv (m_1 , \ldots ,m_S)$, as
\begin{equation}
\redx{\rho}{\mathbf{m}}  \equiv \tr_{\mathbf{\mbarsub}} (\rho ),
\label{eq:A.1}
\end{equation}
where the ``check'' in $\redx{\rho}{\mathbf{m}}$ indicates that it is a \textit{reduction} of \textit{parent state} $\rho$ (and not merely an isolated system of same size as mode group $\mathbf{m}$), and the bar in $\mathbf{\mbar}$ means ``not $\mathbf{m}$,'' telling us to trace over all modes whose labels are \textit{not} in $\mathbf{m}$. See \hyperlink{HedE}{\cite*[App.{\kern 2.5pt}B]{HedE}} for details.
\vspace{-4pt}
\section{\label{sec:App.B}$N$-Separability of $N$-Partite States Implies Reconstructability by Smallest Reductions}
\begin{figure}[H]
\centering
\vspace{-12pt}
\setlength{\unitlength}{0.01\linewidth}
\begin{picture}(100,0)
\put(1,27){\hypertarget{Sec:App.B}{}}
\end{picture}
\end{figure}
\vspace{-39pt}
Recalling the $N$-separable states \smash{$\varsigma ^{(1, \ldots ,N)}$} of \Eq{3}, we will show that each mode-$m$ reduction admits a decomposition of the form \smash{$\redx{\varsigma}{m} \equiv \sum\nolimits_j {p_j \redxshift{\varsigma}{m}{0pt}{0.5pt}_{{\kern -2pt}j} }$} \smash{$=\sum\nolimits_j {p_j \varsigma_{j}^{\shiftmath{0pt}{{\kern 1.5pt}(m)}} }$}, letting us express the parent state entirely in terms of the mode-$m$ decomposition reductions as
\begin{equation}
\varsigma ^{(1, \ldots ,N)} = \sum\nolimits_j {p_j \redx{\varsigma}{1}_j  \otimes  \cdots  \otimes \redx{\varsigma}{N}_j } ,
\label{eq:B.1}
\end{equation}
since \smash{$\redx{\varsigma}{m}_j \equiv \tr_{\mbarsub}(\varsigma _j ^{(1, \ldots ,N)})=\varsigma_{j}^{(m)}$}, echoing the earlier observation from \Sec{I} that absence of entanglement correlation implies that reductions contain enough information to fully reconstruct the parent state.

As we now prove, for all \textit{multimode} reductions (reductions involving two or more modes), $N$-separability of the parent state implies that all multimode reductions are also fully separable, since they too inherit the product-form of the optimal decomposition states of the parent.

First, multipartite reductions are generally mixed, as
\begin{equation}
\redx{\varsigma}{\mathbf{m}}\equiv \sum\nolimits_j {p_j \redx{\varsigma}{\mathbf{m}}_j } ,
\label{eq:B.2}
\end{equation}
where \smash{$\mathbf{m}\!\equiv\! (m_1,\ldots,m_S)$},{\kern -1pt} $S\!\in\! 1,\ldots,N$,{\kern -1pt} and{\kern -1pt} \smash{$\{p_j , \redx{\varsigma}{\mathbf{m}}_j {\kern -1pt}\}$}{\kern -1pt} does not assume any structural similarity to the parent state. Then, from the definition of multipartite reduction,
\begin{equation}
\begin{array}{*{20}l}
   {\redx{\varsigma}{\mathbf{m}} } &\!\! { \equiv \tr_{\mathbf{\mbarsub}} (\varsigma ^{(1, \ldots ,N)} )}  \\
   {\sum\nolimits_j {p_j \redx{\varsigma}{\mathbf{m}}_j } } &\!\! { = \tr_{\mathbf{\mbarsub}} (\sum\nolimits_j {p_j } \varsigma _j ^{(1, \ldots ,N)} )}  \\
   {} &\!\! { = \sum\nolimits_j {p_j \tr_{\mathbf{\mbarsub}} (} \varsigma _j ^{(1, \ldots ,N)} )}  \\
   {} &\!\! { = \sum\nolimits_j {p_j \tr_{\mathbf{\mbarsub}} (} \varsigma _j ^{(1)}  \otimes  \cdots  \otimes \varsigma _j ^{(N)} )}  \\
   {} &\!\! { = \sum\nolimits_j {p_j } \varsigma _j ^{(m_1 )}  \otimes  \cdots  \otimes \varsigma _j ^{(m_S )}, }  \\
\end{array}
\label{eq:B.3}
\end{equation}
where we used the facts that \smash{$\tr(A \otimes B) = \tr(A) \otimes \tr(B)$} \smash{$ = \tr(A)\tr(B)$} and \smash{$\tr(\varsigma_{\shiftmath{0.5pt}{j}}^{{\kern 2.3pt}\shiftmath{-1pt}{({\kern -0.5pt}\mbarsub_k)}})=1$} for normalized states \smash{$\varsigma_{\shiftmath{0.5pt}{j}}^{{\kern 2.3pt}\shiftmath{-1pt}{({\kern -0.5pt}\mbarsub_k)}}$}, which were applicable due to the $N$-separability of the parent.  Setting $S=1$ in \Eq{B.3} yields our earlier result that \smash{$\redxshift{\varsigma}{m}{0pt}{0.5pt}_{{\kern -2pt}j} = \varsigma_{j}^{\shiftmath{0pt}{{\kern 1.5pt}(m)}}$} (since for $S=1$, $\mathbf{m}=m_1=m$), which holds for $m\in 1,\ldots,N$ and lets us rewrite \Eq{B.3} as
\begin{equation}
\redx{\varsigma}{\mathbf{m}}=\sum\nolimits_j {p_j } \redx{\varsigma}{m_{1}}_j  \otimes  \cdots  \otimes \redx{\varsigma}{m_{S}}_j ,
\label{eq:B.4}
\end{equation}
showing that \textit{all} multimode reductions of $N$-separable states are $S$-separable $S$-partite states, so they have no entanglement correlation at all, and can \textit{all} be reconstructed by information in the \textit{single}-mode reductions.  Thus, setting $S=N$ in $\mathbf{m}$ in \Eq{B.4} proves \Eq{B.1} as well.
\section{\label{sec:App.C}Definition of Partitions}
\begin{figure}[H]
\centering
\vspace{-12pt}
\setlength{\unitlength}{0.01\linewidth}
\begin{picture}(100,0)
\put(1,26){\hypertarget{Sec:App.C}{}}
\end{picture}
\end{figure}
\vspace{-39pt}
\textit{Partitioning} is the act of defining \textit{new modes}.  We let the new mode structure of $T$ partitions of multimode reduction $\redx{\rho}{\mathbf{m}}$ be {$\mathbf{m}^{(\mathbf{T})} \equiv (\mathbf{m}^{(1)} | \ldots |\mathbf{m}^{(T)} )$}, where $T\in 1,\ldots,S$ and $\mathbf{m}\equiv(m_{1},\ldots,m_{S})$, and $S\in 1,\ldots,N$, where $\rho$ is an $N$-partite state, and the $T$ new modes defined by the partitioning have internal structures \smash{$\mathbf{m}^{(q)}  \equiv (m_{\shiftmath{-0.5pt}{\scalemath{0.9}{1}}}^{\shiftmath{-0.0pt}{(q)}} , \ldots ,m_{\shiftmath{-0.5pt}{\scalemath{0.8}{G}}}^{\shiftmath{-0.0pt}{(q)}} )$} where $G \equiv G^{(q)}\in 1,\ldots,S$ in terms of original indivisible modes $m_k$ such that all $m_k$ appear exactly once among all new mode groups $\mathbf{m}^{(q)}$.  

The new mode groups have levels vector \smash{$\mathbf{n}^{\shiftmath{-1pt}{(\mathbf{m}^{\shiftmath{-1pt}{(\mathbf{T})}})}}\equiv$} \smash{$(n_{\mathbf{m}^{(1)}},\ldots,n_{\mathbf{m}^{(T)}})$} where \smash{$n_{\mathbf{m}^{(q)}}\!\equiv$}{$\,n$}\smash{\raisebox{3pt}{$\rule{0pt}{4pt}_{m_1^{(q)} }$}}{$ \cdots n$}\smash{\raisebox{3pt}{$\rule{0pt}{4pt}_{m_G^{(q)} }$}}.  Thus we will always have the same number of levels in both \smash{$\redx{\rho}{\mathbf{m}}$} and its partitioned version \smash{$\redxshift{\rho}{\mathbf{m}^{\shiftmath{-1pt}{(\mathbf{T})}}}{-1pt}{0pt}$} so that \smash{$n_{\mathbf{m}}=n_{\mathbf{m}^{(\mathbf{T})}}$}, where \smash{$n_{\mathbf{m}}\equiv n_{m_{1}}\cdots n_{m_{S}}$} is the number of levels of \smash{$\redxshift{\rho}{\mathbf{m}}{-1pt}{0pt}$}, and{\kern 1pt} \smash{$n$}\smash{\raisebox{1pt}{$\rule{0pt}{4pt}_{\mathbf{m}^{(\mathbf{T})}}$}}{$\equiv\!\! n$}\smash{\raisebox{1pt}{$_{\mathbf{m}^{(1)}}\cdots n_{\mathbf{m}^{(T)}}$}} is the number of levels of \smash{$\redxshift{\rho}{\mathbf{m}^{\shiftmath{-1pt}{(\mathbf{T})}}}{-1pt}{0pt}$}. Sometimes we use the notation \smash{$n^{\shiftmath{-1pt}{(\mathbf{m}^{\shiftmath{-1pt}{(\mathbf{T})}})}}\equiv n_{\mathbf{m}^{(\mathbf{T})}}$} to allow space for quantities like \smash{$n_{\shiftmath{0pt}{{\kern 0.5pt}\text{max}}}^{{\kern -0.5pt}\shiftmath{0.5pt}{(\mathbf{m}^{\shiftmath{-1.5pt}{(\mathbf{T})}})}}$} as in \cite{HedE}.

Here, the partition symbol ``$|$'' denotes our conceptual redefinition of the mode structure, so that ``$|$'' is the delimiter of the new mode list, while the commas ``,'' serve as secondary delimiters to be ignored with respect to separability, but shown to indicate how the old modes contribute to the new modes.  In a mode list with no partitions ``$|$'', commas ``,'' \textit{are} the partitions.  Note that we can never subdivide the smallest modes defining the $N$-partite system, since they are defined by the fundamental coincidence behavior of the system and therefore have no internal coincidences of their own (see \hyperlink{HedE}{\cite*[App.{\kern 2.5pt}A]{HedE}}).

Also note that $T$ is the number of mode groups \textit{formed} by the partitions, and is always one more than the number of partition symbols ``$|$''.  For ease of speech, we will speak of ``$T$ partitions'' or describe something as ``$T$-partitional'' when we are referring to it having \textit{$T$ mode groups}, and it is implied that there are always $T-1$ conceptual partitions ``$|$'' that define these mode groups.
\section{\label{sec:App.D}Proof that $N$-Partite States are Entangled If and Only If they are $\text{F}_N$-Partite Entangled}
\begin{figure}[H]
\centering
\vspace{-12pt}
\setlength{\unitlength}{0.01\linewidth}
\begin{picture}(100,0)
\put(1,33){\hypertarget{Sec:App.D}{}}
\end{picture}
\end{figure}
\vspace{-39pt}
First, we establish some useful definitions and a theorem, then we prove necessity and sufficiency separately, in terms of separability, and finally unite the cases.  See \App{A} and \App{C} for supporting explanations.
\vspace{-6pt}
\subsection{\label{sec:App.D.1}Definitions and Theorem}
\hypertarget{Def:1}{\textbf{Definition 1:~~}}Let the set of all unique multimode $T$-partitions of all $S$-mode reductions of an $N$-partite system, where $2\!\leq \!T\!\leq \!S\!\leq \!N$, be called \textit{the set of all multimode $T$-partitions}.  For example: the multimode $T$-partitions of $\rho^{(1,2,3)}$ are $\{\redx{\rho}{1|2},\redx{\rho}{1|3},\redx{\rho}{2|3},\redx{\rho}{1|2,3},$ $\redx{\rho}{2|1,3},\redx{\rho}{3|1,2},\redx{\rho}{1|2|3}\}$.

\hypertarget{Def:2}{\textbf{Definition 2:~~}}Let the $T\geq 2$ modes of a $T$-partitioned state that is not $T$-separable (meaning it cannot be expressed as a convex sum of a tensor product of $T$ pure states) be called \textit{entanglement-correlated} (or \textit{entangled}), and said to have \textit{entanglement correlations} (or \textit{entanglement}) because knowledge of the $T$ single-mode reductions cannot be used to reconstruct the full $T$-mode state.  For example, given parent $\rho\equiv{\kern -1pt}\rho^{(1,2,3)}$, if \smash{$\redx{\rho}{1|3}\neq\sum\nolimits_{\shiftmath{0.5pt}{j}} p_{j}\rho_{{\kern -0.0pt}\shiftmath{1pt}{j}}^{{\kern 2.8pt}\shiftmath{0pt}{(1)}}\otimes\rho_{{\kern -0.0pt}\shiftmath{1pt}{j}}^{{\kern 2.8pt}\shiftmath{0pt}{(3)}}$} for some pure \smash{$\rho_{{\kern -0.0pt}\shiftmath{1pt}{j}}^{{\kern 2.8pt}\shiftmath{0pt}{(1)}}$} and \smash{$\rho_{{\kern -0.0pt}\shiftmath{1pt}{j}}^{{\kern 2.8pt}\shiftmath{0pt}{(3)}}$}, then modes $1$ and $3$ of $\rho$ are entanglement-correlated, so the reduction $\redx{\rho}{1,3}=\redx{\rho}{1|3}$ has entanglement correlations.\hsp{-0.3} (We say \textit{entanglement correlation} instead of just \textit{entanglement} as a reminder that there are \textit{other} types of nonlocal correlation that do not involve entanglement.)

\hypertarget{Thm:1}{\textbf{Theorem\hsp{-0.6} 1:~~}}\hsp{-3}\textit{The\hsp{-0.6} set\hsp{-0.6} of\hsp{-0.6} all\hsp{-0.6} multimode\hsp{-0.6} $T$-partitions\hsp{-0.6} from\hsp{-0.6} \hyperlink{Def:1}{Definition\hsp{-0.6} 1}\hsp{-0.6} is\hsp{-0.6} the\hsp{-0.6} set\hsp{-0.6} all\hsp{-0.6} possible\hsp{-0.6} mode\hsp{-0.6} groups\hsp{-0.6} that\hsp{-0.6} could\hsp{-0.6} exhibit\hsp{-0.6} entanglement\hsp{-0.6} correlations\hsp{-0.6} within\hsp{-0.6} a\hsp{-0.6} state}.\hsp{-0.6} Proof:\hsp{-0.6} (i)\hsp{-0.6} It\hsp{-0.6} is\hsp{-0.6} not\hsp{-0.6} possible\hsp{-0.6} to\hsp{-0.6} define\hsp{-0.6} any\hsp{-0.6} other\hsp{-0.6} multimode\hsp{-0.6} groups\hsp{-0.6} within\hsp{-0.6} the\hsp{-0.6} system,\hsp{-0.6} since\hsp{-0.6} the\hsp{-0.6} original\hsp{-0.6} modes\hsp{-0.6} cannot\hsp{-0.6} be\hsp{-0.6} subdivided\hsp{-0.6} (see \hyperlink{HedE}{\cite*[App.{\kern 2.5pt}A]{HedE}});\hsp{-0.6} therefore\hsp{-0.6} this\hsp{-0.6} list\hsp{-0.6} of\hsp{-0.6} mode\hsp{-0.6} groups\hsp{-0.6} is\hsp{-0.6} exhaustive.\hsp{-0.6} (ii)\hsp{-0.6} By\hsp{-0.6} \hyperlink{Def:2}{Definition\hsp{-0.6} 2},\hsp{-0.6} entanglement\hsp{-0.6} correlations\hsp{-0.6} can\hsp{-0.6} only\hsp{-0.6} exist\hsp{-0.6} between\hsp{-0.6} two\hsp{-0.6} or\hsp{-0.6} more\hsp{-0.6} modes.\hsp{-0.6}  (iii)\hsp{-0.6} Therefore,\hsp{-0.6} by\hsp{-0.6} (i)\hsp{-0.6} and\hsp{-0.6} (ii),\hsp{-0.6} \hyperlink{Thm:1}{Theorem\hsp{-0.6} 1}\hsp{-0.6} is\hsp{-0.6} proven.

For applications of entanglement in the full Hilbert space (not within reductions), we will use a relaxed version of \hyperlink{Thm:1}{Theorem 1}, that only uses the set of multimode $T$-partitions of the full $N$-mode state, ignoring its reductions. See \Sec{III.D}.
\vspace{-6pt}
\subsection{\label{sec:App.D.2}Proof That $N$-Separability Is Sufficient for Absence of All Entanglement Correlations}
We already proved in \Eq{B.4} that all $S$-mode reductions $\redx{\varsigma}{\mathbf{m}}$ of $N$-separable $N$-partite states $\varsigma$ are $S$-separable.  Therefore, by \hyperlink{Thm:1}{Theorem 1}, in order to show that all possible entanglement correlations of $\varsigma$ are absent, we need to show for any \textit{partitions} of $\mathbf{m}\equiv(m_{1},\ldots,m_{S})$ in the states of \Eq{B.4} such as $\mathbf{m}'\equiv(m_{1}|m_{2},m_{3})$, that such states are \textit{also separable across all partitions}, a fact easily proven since partitions in $S$-separable states only result in selectively ignoring separability between certain modes.

For an example of how $T$-partite partitioning of $S$-separable $S$-partite states yields $T$-separability, observe that $2$-partition of a $3$-partite $3$-separable state gives {$\redx{\varsigma}{\mathbf{m}'}\equiv\redx{\varsigma}{m_{1}|m_{2},m_{3}}=\sum\nolimits_{\shiftmath{1pt}{j}} {p_j } \redxshift{\varsigma}{m_{1}}{0pt}{1.5pt}_{{\kern -1.5pt}\shiftmath{0.7pt}{j}}  \otimes  (\redxshift{\varsigma}{m_{2}}{0pt}{1.5pt}_{{\kern -1.5pt}\shiftmath{0.7pt}{j}}  \otimes \redxshift{\varsigma}{m_{3}}{0pt}{1.5pt}_{{\kern -1.5pt}\shiftmath{0.7pt}{j}})\equiv$} \smash{$\sum\nolimits_{\shiftmath{1pt}{j}} {p_j } \redxshift{\varsigma}{m_{1}}{0pt}{1.5pt}_{{\kern -1.5pt}\shiftmath{0.7pt}{j}}  \otimes  \redxshift{\varsigma}{m_{2},\!m_{3}}{0pt}{1.5pt}_{{\kern -1.5pt}\shiftmath{0.7pt}{j}}$},{\kern -1pt} where \smash{$\redxshift{\varsigma}{m_{2},\!m_{3}}{0pt}{1.5pt}_{{\kern -1.5pt}\shiftmath{0.7pt}{j}}\!\equiv\!\redxshift{\varsigma}{m_{2}}{0pt}{1.5pt}_{{\kern -1.5pt}\shiftmath{0.7pt}{j}}  \!\otimes \redxshift{\varsigma}{m_{3}}{0pt}{1.5pt}_{{\kern -1.5pt}\shiftmath{0.7pt}{j}}$}, showing that in $S$-separable states, partitioning means grouping modes together and ignoring their internal separability, so \textit{the resulting mode groups are separable with each other}, due to the underlying $S$-separability of \smash{$\redx{\varsigma}{\mathbf{m}}$}.

For a general proof of this, let $\sigma$ be a $T$-separable $T$-partitioned $S$-mode mixed state with the form
\begin{equation}
\sigma\equiv \sigma^{(\mathbf{m}^{(1)} | \ldots |\mathbf{m}^{(T)} )}\equiv \sum\nolimits_j {p_j \sigma_j^{(\mathbf{m}^{(1)})}   \otimes  \cdots  \otimes \sigma_j^{(\mathbf{m}^{(T)})}  } ,
\label{eq:D.1}
\end{equation}
where \smash{$\sigma_j^{(\mathbf{m}^{(q)})}$} are new-mode-\smash{$\mathbf{m}^{(q)}$} pure states of the optimal decomposition, keeping in mind that for a given $T$, there is generally more than one way to partition the state to $T$ new modes.  

Then, similarly to \Eq{B.3} and \Eq{B.4}, the \smash{$\mathbf{m}^{(\mathbf{q})}$}-mode reductions \smash{$\redxshift{\sigma{\kern -1.1pt}}{\mathbf{m}^{\shiftmath{-1pt}{(\mathbf{q})}}}{-1pt}{0pt}\equiv\sum\nolimits_{j}p_{j}\redxshift{\sigma{\kern -1.1pt}}{\mathbf{m}^{\shiftmath{-1pt}{(\mathbf{q})}}}{0pt}{1.5pt}_{{\kern -1.5pt}\shiftmath{0.7pt}{j}}$} of $\sigma$, where $\mathbf{q}\equiv(q_{1},\ldots,q_{Q})$ and $Q\in 1,\ldots,\scalemath{0.98}{T}$, are
\begin{equation}
\begin{array}{*{20}l}
   {\redx{\sigma{\kern -1.1pt}}{\mathbf{m}^{(\mathbf{q})}}} &\!\! { \equiv \tr_{{\kern 1.5pt}\scalemath{1}{\shiftmath{-0.3pt}{\overline{\rule{0pt}{6pt}{\kern -0.1pt}\mathbf{m}^{\shiftmath{-0.5pt}{(\mathbf{q})}}}}}{\kern 0.5pt}} (\sigma^{(\mathbf{m}^{(1)} | \ldots |\mathbf{m}^{(T)} )})}  \\
   {\sum\nolimits_j {p_j \redx{\sigma{\kern -1.1pt}}{\mathbf{m}^{(\mathbf{q})}}_j } } &\!\! { = \sum\nolimits_j {p_j}\tr_{{\kern 1.5pt}\scalemath{1}{\shiftmath{-0.3pt}{\overline{\rule{0pt}{6pt}{\kern -0.1pt}\mathbf{m}^{\shiftmath{-0.5pt}{(\mathbf{q})}}}}}{\kern 0.5pt}}( \sigma_j^{(\mathbf{m}^{(1)})}   \otimes  \cdots  \otimes \sigma_j^{(\mathbf{m}^{(T)})} )}  \\
   {} &\!\! { = \sum\nolimits_j {p_j } \sigma^{(\mathbf{m}^{(q_{1})})}_{j}  \otimes  \cdots  \otimes \sigma^{(\mathbf{m}^{(q_{Q})})}_{j}, }  \\
\end{array}
\label{eq:D.2}
\end{equation}
and then, the $Q=1$ case shows that \smash{$\redxshift{\sigma{\kern -1.1pt}}{\mathbf{m}^{\shiftmath{-1pt}{(q)}}}{-0.5pt}{2pt}_{{\kern -1.5pt}\shiftmath{0.7pt}{j}}=\sigma^{{\kern 2.5pt}\shiftmath{-0.5pt}{(\mathbf{m}^{\shiftmath{-1pt}{(q)}})}}_{{\kern -0.5pt}\shiftmath{0.7pt}{j}}$} (since then $\mathbf{q}=q_{1}\equiv q$), which, put into \Eq{D.2}, yields
\begin{equation}
\redx{\sigma{\kern -1.1pt}}{\mathbf{m}^{(\mathbf{q})}}=\sum\nolimits_j {p_j } \redx{\sigma{\kern -1.1pt}}{\mathbf{m}^{(q_{1})}}_{j}  \otimes  \cdots  \otimes \redx{\sigma{\kern -1.1pt}}{\mathbf{m}^{(q_{Q})}}_{j},
\label{eq:D.3}
\end{equation}
and then the $Q=T$ case of \Eq{D.3} yields the useful result
\begin{equation}
\redx{\sigma{\kern -1.1pt}}{\mathbf{m}^{(\mathbf{T})}}=\sum\nolimits_j {p_j } \redx{\sigma{\kern -1.1pt}}{\mathbf{m}^{(1)}}_{j}  \otimes  \cdots  \otimes \redx{\sigma{\kern -1.1pt}}{\mathbf{m}^{(T)}}_{j},
\label{eq:D.4}
\end{equation}
since we can choose $\mathbf{q}=(q_{1},\ldots,q_{T})=(1,\ldots,T)=\mathbf{T}$, which shows that a $T$-separable $T$-partite state $\sigma$ can be fully described by information in its smallest reductions \smash{$\redxshift{\sigma{\kern -1.1pt}}{\mathbf{m}^{\shiftmath{-1pt}{(q)}}}{-1pt}{0pt}\equiv\sum\nolimits_{{\kern 1.3pt}\shiftmath{1.5pt}{j}}{\kern -0.5pt}p_{j}\redxshift{\sigma{\kern -1.1pt}}{\mathbf{m}^{\shiftmath{-1pt}{(q)}}}{0pt}{1.5pt}_{{\kern -1.5pt}\shiftmath{0.7pt}{j}}$}\!, \textit{even if those reductions have internal{\kern -0.9pt} mode{\kern -0.9pt} structure}{\kern -0.9pt} \smash{$\mathbf{m}^{(q)}  \equiv (m_{\shiftmath{-0.5pt}{\scalemath{0.9}{1}}}^{\shiftmath{-0.0pt}{(q)}} , \ldots ,m_{\shiftmath{-0.5pt}{\scalemath{0.8}{G}}}^{\shiftmath{-0.0pt}{(q)}} )$}.  

Then, the key point is that since none of the original modes $m_k$ of an $N$-separable $N$-partite state $\varsigma$ can be subdivided by partitions, the \smash{$\mathbf{m}^{\shiftmath{-0.5pt}{(\mathbf{T})}}$}-mode reductions of $\varsigma$ are guaranteed to be $T$-separable, as
\begin{equation}
\begin{array}{*{20}l}
   {\tr_{{\kern 1.5pt}\scalemath{1}{\shiftmath{-0.3pt}{\overline{\rule{0pt}{6pt}{\kern -0.1pt}\mathbf{m}^{\shiftmath{-0.5pt}{(\mathbf{T})}}}}}{\kern 0.5pt}} (\varsigma)} &\!\! {=\tr_{{\kern 1.5pt}\scalemath{1}{\shiftmath{-0.3pt}{\overline{\rule{0pt}{6pt}{\kern -0.1pt}\mathbf{m}^{\shiftmath{-0.5pt}{(\mathbf{T})}}}}}{\kern 0.5pt}} (\varsigma^{(1,\ldots,N )})}  \\
   {\redx{\varsigma}{\mathbf{m}^{(\mathbf{T})}}} &\!\! {=\sum\nolimits_j {p_j \tr_{{\kern 1.5pt}\scalemath{1}{\shiftmath{-0.3pt}{\overline{\rule{0pt}{6pt}{\kern -0.1pt}\mathbf{m}^{\shiftmath{-0.5pt}{(\mathbf{T})}}}}}{\kern 0.5pt}} }(\redx{\varsigma}{1}_{j}\otimes\cdots\otimes\redx{\varsigma}{N}_{j})}  \\
   {} &\!\! {=\sum\nolimits_j {p_j \redxshift{\varsigma}{\mathbf{m}^{\shiftmath{-1pt}{(1)}}}{-0.5pt}{2pt}_{{\kern -1pt}j}\otimes\cdots\otimes\redxshift{\varsigma}{\mathbf{m}^{\shiftmath{-1pt}{(T)}}}{-0.5pt}{2pt}_{{\kern -1pt}j} },}  \\
\end{array}
\label{eq:D.5}
\end{equation}
where \smash{$\redxshift{\varsigma}{\mathbf{m}^{\shiftmath{-1pt}{(q)}}}{-0.5pt}{2pt}_{{\kern -1.5pt}\shiftmath{0.7pt}{j}}\equiv\redxshift{\varsigma}{m_{{\kern -0.5pt}\shiftmath{-1.0pt}{\scalemath{0.9}{1}}}^{\shiftmath{-0.5pt}{(q)}}}{-0.5pt}{2pt}_{{\kern -1.5pt}\shiftmath{0.7pt}{j}}\otimes\cdots\otimes\redxshift{\varsigma}{m_{{\kern -0.5pt}\shiftmath{-1.0pt}{\scalemath{0.9}{G}}}^{\shiftmath{-0.5pt}{(q)}}}{-0.5pt}{2pt}_{{\kern -1.5pt}\shiftmath{0.7pt}{j}}$} are pure decomposition{\kern 0.8pt} states{\kern 0.8pt} of{\kern 0.8pt} \smash{$\redxshift{\varsigma}{\mathbf{m}^{\shiftmath{-1pt}{(q)}}}{-0.5pt}{0pt}$}.

Thus we have proven that if an $N$-partite parent state is $N$-separable, all of its multimode reductions to $S\in 2,\ldots,N$ modes are $S$-separable $S$-partite states, and any \textit{partitions} of any $S$-partite reductions of $N$-separable $N$-partite states for $S\in 2,\ldots,N$, are \textit{also} separable across those partitions.  Therefore, \textit{$N$-separability of $N$-partite states implies that there are no entanglement correlations of any kind}.  This yields the equivalent statements,
\begin{equation}
\begin{array}{*{20}c}
   {\begin{array}{*{20}c}
   {\left({{\kern 0.5pt}\mbox{$N$-separability of the full state}{\kern 1pt}}\right)}  \\
   {\mbox{is sufficient for}}  \\
   {\,\left({
\begin{array}{*{20}c}
   {\mbox{the simultaneous absence of}}  \\
   {\mbox{all entanglement correlations}}  \\
\end{array}
}\right)}  \\
\end{array}} & {\,(\text{S} \Leftarrow \text{N}),}  \\
\end{array}
\label{eq:D.6}
\end{equation}
\begin{equation}
\begin{array}{*{20}c}
   {\begin{array}{*{20}c}
   {\left(\hsp{0.5}{\mbox{$\text{F}_N$-entanglement of the full state}}{\kern 1pt}\right)}  \\
   {\mbox{is necessary for}}  \\
   {\,\left({
\begin{array}{*{20}c}
   {\mbox{the presence of any}}  \\
   {\mbox{entanglement correlations}}  \\
\end{array}
}\right)}  \\
\end{array}} &\!\!\!\! {\,(\overline{\text{S}\rule{0pt}{9pt}} \Rightarrow \overline{\text{N}\rule{0pt}{9pt}}),}  \\
\end{array}
\label{eq:D.7}
\end{equation}
where $\text{S}$ and $\text{N}$ are labels for conditions of a conditional statement where we chose $\text{N}$ to represent ``$N$-separability of the full state'' and $\text{S}$ to represent ``the simultaneous absence of all entanglement correlations,'' and the phrase ``the full state'' means ``the full $N$-partite parent state.''
\subsection{\label{sec:App.D.3}Proof That $N$-Separability Is Necessary for Absence of All Entanglement Correlations}
Here, the claim we want to test is
\begin{equation}
\begin{array}{*{20}c}
   {\begin{array}{*{20}c}
   {\left({\kern 0.5pt}{\mbox{$N$-separability of the full state}}{\kern 1pt}\right)}  \\
   {\mbox{is necessary for}}  \\
   {\,\left({
\begin{array}{*{20}c}
   {\mbox{the simultaneous absence of}}  \\
   {\mbox{all entanglement correlations}}  \\
\end{array}
}\right)}  \\
\end{array}} &\!\!\!\! {\,(\text{S}\Rightarrow \text{N}).}  \\
\end{array}
\label{eq:D.8}
\end{equation}
To prove \Eq{D.8}, suppose that $N$-separability is \textit{not} necessary for the simultaneous absence of all entanglement correlations. Then, that means there could exist states $\varrho$ that could be $\text{F}_N$-entangled, and yet \textit{also} have simultaneous absence of all entanglement correlations.  Therefore, the $\text{F}_N$-entanglement of such states would mean that
\begin{equation}
\varrho ^{(1, \ldots ,N)} \neq \sum\nolimits_j {p_j \redx{\varrho}{1}_j  \otimes  \cdots  \otimes \redx{\varrho}{N}_j } ,
\label{eq:D.9}
\end{equation}
while their simultaneous absence of all entanglement correlations means that all of their $T$-partitioned $S$-mode reductions must be $T$-separable,\hsp{-0.4} so\hsp{-0.4} that,\hsp{-0.4} as\hsp{-0.4} proved\hsp{-0.4} in\hsp{-0.4} \Eq{D.5},
\begin{equation}
\redx{\varrho}{\mathbf{m}^{(\mathbf{T})}}=\sum\nolimits_j {p_j } \redx{\varrho}{\mathbf{m}^{(1)}}_{j}  \otimes  \cdots  \otimes \redx{\varrho}{\mathbf{m}^{(T)}}_{j},
\label{eq:D.10}
\end{equation}
for all $2\leq T\leq S$ and $2\leq S \leq N$, where $S$ is the number of modes of a reduction without the partitions. Then, computing all $T$-partitioned $S$-mode reductions of \Eq{D.9} by taking the partial trace gives
\begin{equation}
\tr_{{\kern 1.5pt}\scalemath{1}{\shiftmath{-0.3pt}{\overline{\rule{0pt}{6pt}{\kern -0.1pt}\mathbf{m}^{\shiftmath{-0.5pt}{(\mathbf{T})}}}}}{\kern 0.5pt}} (\varrho^{(1,\ldots,N)})\neq\sum\nolimits_j {p_j } \redx{\varrho}{\mathbf{m}^{(1)}}_{j}  \otimes  \cdots  \otimes \redx{\varrho}{\mathbf{m}^{(T)}}_{j},
\label{eq:D.11}
\end{equation}
but since \smash{$\redx{\varrho}{\mathbf{m}^{(\mathbf{T})}}\equiv\tr_{{\kern 1.5pt}\scalemath{1}{\shiftmath{-0.3pt}{\overline{\rule{0pt}{6pt}{\kern -0.1pt}\mathbf{m}^{\shiftmath{-0.5pt}{(\mathbf{T})}}}}}{\kern 0.5pt}} (\varrho^{(1,\ldots,N)})$} by definition, then we can put \Eq{D.10} into the left side of \Eq{D.11} to get
\begin{equation}
\sum\nolimits_{{\kern -0.5pt}j} {\!{\kern -0.0pt}p_j } \redx{\varrho}{\mathbf{m}^{(1)}{\kern -0.3pt}}_{j}  {\kern -1.0pt}{\otimes  \cdots  \otimes}{\kern -0.2pt} \redx{\varrho}{\mathbf{m}^{(T)}{\kern -0.3pt}}_{j}\!{\kern -1.5pt}\neq{\kern -4.5pt}\sum\nolimits_{{\kern -0.5pt}j} {\!{\kern -0.0pt}p_j } \redx{\varrho}{\mathbf{m}^{(1)}{\kern -0.3pt}}_{j}  {\kern -1.0pt}{\otimes  \cdots  \otimes}{\kern -0.2pt} \redx{\varrho}{\mathbf{m}^{(T)}{\kern -0.3pt}}_{j}\!\!,
\label{eq:D.12}
\end{equation}
which is a \textit{false statement}, meaning that the supposition is false.  Thus, the statement in \Eq{D.8} is \textit{true}, and we can extract from it the corresponding statement that
\begin{equation}
\begin{array}{*{20}c}
   {\begin{array}{*{20}c}
   {\left({\kern 0.5pt}{\mbox{$\text{F}_N$-entanglement of the full state}}{\kern 1pt}\right)}  \\
   {\mbox{is sufficient for}}  \\
   {\,\left({
\begin{array}{*{20}c}
   {\mbox{the presence of any}}  \\
   {\mbox{entanglement correlations}}  \\
\end{array}
}\right)}  \\
\end{array}} &\!\!\!\! {\,(\overline{\text{S}\rule{0pt}{9pt}} \Leftarrow \overline{\text{N}\rule{0pt}{9pt}}).}  \\
\end{array}
\label{eq:D.13}
\end{equation}
\subsection{\label{sec:App.D.4}Unification of Results}
Together, true statements \Eq{D.6} and \Eq{D.8} yield
\begin{equation}
\begin{array}{*{20}c}
   {\begin{array}{*{20}c}
   {\left({{\kern 0.5pt}\mbox{$N$-separability of the full state}{\kern 1pt}}\right)}  \\
   {\mbox{is necessary and sufficient for}}  \\
   {\,\left({
\begin{array}{*{20}c}
   {\mbox{the simultaneous absence of}}  \\
   {\mbox{all entanglement correlations}}  \\
\end{array}
}\right)}  \\
\end{array}} &\hsp{-3} {\,(\text{S} \Leftrightarrow \text{N}),}  \\
\end{array}
\label{eq:D.14}
\end{equation}
which means we also have
\begin{equation}
\begin{array}{*{20}c}
   {\begin{array}{*{20}c}
   {\left({\kern 0.5pt}{\mbox{$\text{F}_N$-entanglement of the full state}}{\kern 1pt}\right)}  \\
   {\mbox{is necessary and sufficient for}}  \\
   {\,\left({
\begin{array}{*{20}c}
   {\mbox{the presence of any}}  \\
   {\mbox{entanglement correlations}}  \\
\end{array}
}\right)}  \\
\end{array}} &\!\!\!\! {\,(\overline{\text{S}\rule{0pt}{9pt}} \Leftrightarrow \overline{\text{N}\rule{0pt}{9pt}}).}  \\
\end{array}
\label{eq:D.15}
\end{equation}
Thus,\hsp{-0.5} we\hsp{-0.5} have\hsp{-0.5} proven\hsp{-0.5} that\hsp{-0.5} $\text{F}_N$-entanglement\hsp{-0.5} measures\hsp{-0.5} are\hsp{-0.5} necessary\hsp{-0.5} and\hsp{-0.5} sufficient\hsp{-0.5} for\hsp{-0.5} detecting\hsp{-0.5} the\hsp{-0.5} presence\hsp{-0.5} of\hsp{-0.5} any\hsp{-0.5} entanglement\hsp{-0.5} correlations\hsp{-0.5} in\hsp{-0.5} an\hsp{-0.5} $N$-partite\hsp{-0.5} quantum\hsp{-0.5} state.
\section{\label{sec:App.E}Normalization Factor of the Ent}
\begin{figure}[H]
\centering
\vspace{-12pt}
\setlength{\unitlength}{0.01\linewidth}
\begin{picture}(100,0)
\put(1,23){\hypertarget{Sec:App.E}{}}
\end{picture}
\end{figure}
\vspace{-39pt}
Given the parameters from \Eq{5}, the ent's automatic normalization (needing no calibration state) function is
\begin{equation}
M(L) \equiv M(L,\mathbf{n})\equiv 1 - \frac{1}{N}\sum\limits_{m = 1}^N {\frac{{n_m P_{\text{MP}}^{(m)}(L) - 1}}{{n_m  - 1}}}\,,
\label{eq:E.1}
\end{equation}
where $N\equiv\text{dim}(\mathbf{n})$ and the mode-$m$ purity-minimizing function \smash{$P_{\text{MP}}^{(m)} (L)\equiv P_{\text{MP}}^{(m)} (L,\mathbf{n})$} is
\begin{equation}
\begin{array}{*{20}l}
   {P_{\text{MP}}^{(m)} (L,\mathbf{n})\equiv } &\!\! {\text{mod} (L ,n_m )\left( {\frac{{1 + \text{floor}(L /n_m )}}{{L }}} \right)^2 }  \\
   {} &\!\! { + (n_m  - \text{mod} (L ,n_m ))\left( {\frac{{\text{floor}(L /n_m )}}{{L }}} \right)^2 ,}  \\
\end{array}
\label{eq:E.2}
\end{equation}
where \smash{$\text{mod}(a,b) \equiv a - \text{floor}(\frac{a}{b})b$}.  The minimum physical purity of \smash{$\redx{\rho}{m}$} is then \smash{$P_{\text{MP}}^{\shiftmath{0.5pt}{(m)}} (L_* )$}, where \smash{$L_*\equiv L_*(\mathbf{n})$} is any number of levels of equal nonzero probabilities that can support maximal $\text{F}_N$-entanglement, given by
\begin{equation}
\{ L_* \}\equiv\{ L \} \;\;\text{s.t.}\;\!\mathop {\min }\limits_{L \in 2, \ldots, \nmaxnot } (1 - M(L)),
\label{eq:E.3}
\end{equation}
where \smash{$\nmaxnot \equiv \frac{n}{{n_{\max } }}$} is the product of all $n_m$ \textit{except} $n_{\max }$, where \smash{$n_{\max } \equiv \max (\mathbf{n})$}, \smash{$\mathbf{n} \equiv (n_1 , \ldots ,n_N )$} and $n\equiv n_{1}\cdots n_{N}$.  Thus, $M(L_{*})$ is the factor in \Eq{5}. See \cite{HedE} for details.  For $N$-qu$d$it systems (\smash{$n_{m}=d\;\forall m$}), \smash{$M(L_{*})=1$}.
\section{\label{sec:App.F}True-Generalized X (TGX) States}
\begin{figure}[H]
\centering
\vspace{-12pt}
\setlength{\unitlength}{0.01\linewidth}
\begin{picture}(100,0)
\put(1,24){\hypertarget{Sec:App.F}{}}
\end{picture}
\end{figure}
\vspace{-39pt}
Explained further in \hyperlink{HedE}{\cite*[App.{\kern 2.5pt}D]{HedE}}, and first presented in \cite{HedX}, true-generalized X (TGX) states are a special family of density matrices that are conjectured to be related to all general states (pure and mixed) by an entanglement-preserving unitary (EPU) transformation such that the general state and the TGX state have the same entanglement, a property called \textit{EPU equivalence}.

Restricting ourselves to $\text{F}_N$-entanglement, the most likely candidate for TGX states are \textit{simple} states, defined as those for which all of the off-diagonal parent-state matrix elements appearing in the off-diagonals of the $N$ single-mode reductions are identically zero.

For example, in $2\times 2$, the TGX states are X states,
\begin{equation}
\scalebox{0.95}{$\rho  = \!\left( {\begin{array}{*{20}c}
   {\rho _{1,1} } & \cdot & \cdot & {\rho _{1,4} }  \\
   \cdot & {\rho _{2,2} } & {\rho _{2,3} } & \cdot  \\
   \cdot & {\rho _{3,2} } & {\rho _{3,3} } & \cdot  \\
   {\rho _{4,1} } & \cdot & \cdot & {\rho _{4,4} }  \\
\end{array}} \right)\!,$}
\label{eq:F.1}
\end{equation}
where dots are zeros, while in $2\times 3$, the TGX states are
\begin{equation}
\scalebox{0.95}{$\rho  =\! \left( {\begin{array}{*{20}c}
   {\rho _{1,1} } &  \cdot  &  \cdot  &  \cdot  & {\rho _{1,5} } & {\rho _{1,6} }  \\
    \cdot  & {\rho _{2,2} } &  \cdot  & {\rho _{2,4} } &  \cdot  & {\rho _{2,6} }  \\
    \cdot  &  \cdot  & {\rho _{3,3} } & {\rho _{3,4} } & {\rho _{3,5} } &  \cdot   \\
    \cdot  & {\rho _{4,2} } & {\rho _{4,3} } & {\rho _{4,4} } &  \cdot  &  \cdot   \\
   {\rho _{5,1} } &  \cdot  & {\rho _{5,3} } &  \cdot  & {\rho _{5,5} } &  \cdot   \\
   {\rho _{6,1} } & {\rho _{6,2} } &  \cdot  &  \cdot  &  \cdot  & {\rho _{6,6} }  \\
\end{array}} \right)\!,$}
\label{eq:F.2}
\end{equation}
see \hyperlink{HedE}{\cite*[App.{\kern 2.5pt}D]{HedE}} to see how these were obtained.  Thus, \Eq{F.2} shows that TGX states are not always X states, and \cite{HedX} gave numerical evidence that \Eq{F.2} can reach values of entanglement for certain rank and purity combinations not accessible to X states, which was later proven in \cite{MeMH}, and proves that X states \textit{cannot} exhibit EPU equivalence in general (though they can in some systems), while numerical evidence in both \cite{HedX} and \cite{MeMH} indicates that TGX states \textit{may} indeed have EPU equivalence in $2\times 3$ systems.  Furthermore, the conjecture of EPU equivalence from \cite{HedX} was proven for the $2\times 2$ case in \cite{MeMG}.

In larger multipartite systems, this definition of TGX states from \cite{HedX} led to the discovery of a set of TGX states that were proven in \cite{HedE} to be maximally $\text{F}_N$-entangled, and also led to the proof of the existence of maximally entangled basis (MEB) sets in \cite{HedE}, first conjectured in \cite{HedX}. Thus, the TGX states contain enough maximally $\text{F}_N$-entangled states to form a complete basis in every multipartite system.

Therefore, while the exact form of TGX states is still unproven with respect to their defining property of EPU equivalence, the hypothesis that they are \textit{simple} states as defined above has been shown to be consistent with EPU equivalence in many numerical and analytical tests.
\section{\label{sec:App.G}Full Set of $4$-Qubit Maximally $\text{F}_N$-Entangled TGX States Involving $|1\rangle$}
\begin{figure}[H]
\centering
\vspace{-12pt}
\setlength{\unitlength}{0.01\linewidth}
\begin{picture}(100,0)
\put(1,27){\hypertarget{Sec:App.G}{}}
\end{picture}
\end{figure}
\vspace{-39pt}
From \cite{HedE}, the 13-step algorithm yields the full set of $4$-qubit maximally $\text{F}_N$-entangled TGX states involving $|1\rangle$, with level-label convention $\{|1\rangle,\!|2\rangle,\ldots,|16\rangle\}\equiv\{|1,\!1,\!1,\!1\rangle,|1,\!1,\!1,\!2\rangle,\ldots,|2,\!2,\!2,\!2\rangle\}$, as
\begin{equation}
\begin{array}{*{20}l}
   {|\Phi _j^{[L_* ]} \rangle} &\!\! {\equiv {\textstyle{1 \over {\rule{0pt}{7pt}{\kern -0.1pt}\sqrt {L_* } }}}\sum\nolimits_{k = 1}^{L_* } {|(L_{\text{ME}}^{[L_* ]} )_{j,k} \rangle } ,}  \\
\end{array}
\label{eq:G.1}
\end{equation}
where $L_{*}$ is the number of \textit{nonzero levels} (nonzero probability amplitudes of these states), and\hsp{-1} \smash{$L_{\text{ME}}^{[L_* ]}$} is the matrix of generic-basis-level labels for a given $L_*$, given by
\begin{equation}
\scalebox{0.90}{$L_{\text{ME}}^{[2]}  = \left( {\begin{array}{*{20}c}
   1 & {16}  \\
\end{array}} \right),\;\;L_{\text{ME}}^{[4]}  = \!\left( {\begin{array}{*{20}c}
   1 & 4 & {13} & {16}  \\
   1 & 4 & {14} & {15}  \\
   1 & 6 & {11} & {16}  \\
   1 & 6 & {12} & {15}  \\
   1 & 7 & {10} & {16}  \\
   1 & 7 & {12} & {14}  \\
   1 & 8 & {10} & {15}  \\
   1 & 8 & {11} & {14}  \\
   1 & 8 & {12} & {13}  \\
\end{array}} \right)\!,$}
\label{eq:G.2}
\end{equation}
for $L_{*}=2$ and $L_{*}=4$, while for $L_{*}=6$ and $L_{*}=8$,
\begin{equation}
\scalebox{0.90}{$\begin{array}{*{20}l}
   {L_{\text{ME}}^{[6]} } &\!\! { = \!\left(\! {\begin{array}{*{20}c}
   1 & 4 & 6 & {11} & {13} & {16}  \\
   1 & 4 & 7 & {10} & {13} & {16}  \\
   1 & 6 & 7 & {10} & {11} & {16}  \\
\end{array}} \right)\!,}  \\
   {L_{\text{ME}}^{[8]} } &\!\! { = \left( {\begin{array}{*{20}c}
   1 & 4 & 6 & 7 & {10} & {11} & {13} & {16}  \\
\end{array}} \right).}  \\
\end{array}$}
\label{eq:G.3}
\end{equation}

In \Eq{14}, \smash{$|\Phi _{\text{GHZ}} \rangle  \equiv |\Phi _1^{[2]} \rangle$}, \smash{$|\Phi _{\text{BP} } \rangle  \equiv |\Phi _1^{[4]} \rangle$}, and \smash{$|\Phi _{\text{F}} \rangle  \equiv$} \smash{$ |\Phi _2^{[4]}\rangle$}. We show these both to show what was used in our tests, and because any tests of new measures for four qubits would benefit from starting with this set as well.
\section{\label{sec:App.H}Quantum Mixed States Cannot Be Treated as Time-Averages of Varying Pure States}
\begin{figure}[H]
\centering
\vspace{-12pt}
\setlength{\unitlength}{0.01\linewidth}
\begin{picture}(100,0)
\put(1,27){\hypertarget{Sec:App.H}{}}
\end{picture}
\end{figure}
\vspace{-39pt}
The reason that states such as \Eq{2} are considered biseparable is that ``they can be prepared through a statistical mixture of biparitite [and biseparable] entangled states'' \cite{HSGS,HMGH}.  However, we must be careful not to think of such states as \textit{separable}; while it is true that making a step function of the different biseparable pure states of that decomposition could yield identical tomographic results to an actual quantum mixed state of the same form,\hsp{-0.3} a\hsp{-0.3} mixture\hsp{-0.3} obtained\hsp{-0.3} from\hsp{-0.3} a\hsp{-0.3} time-average\hsp{-0.3} of\hsp{-0.3} a\hsp{-0.3} step\hsp{-0.3} function\hsp{-0.3} of\hsp{-0.3} pure\hsp{-0.3} states\hsp{-0.3} (as\hsp{-0.3} the\hsp{-0.3} term\hsp{-0.3} ``prepare''\hsp{-0.3} may\hsp{-0.3} suggest\hsp{-0.3} to\hsp{-0.3} some)\hsp{-0.3} is\hsp{-0.3} merely\hsp{-0.3} an\hsp{-0.3} \textit{estimation}\hsp{-0.3} resulting\hsp{-0.3} from\hsp{-0.3} the\hsp{-0.3} measurer's\hsp{-0.3} ignorance\hsp{-0.3} about\hsp{-0.3} which\hsp{-0.3} measurements\hsp{-0.3} correspond\hsp{-0.3} to\hsp{-0.3} which\hsp{-0.3} pure\hsp{-0.3} states\hsp{-0.3} of\hsp{-0.3} the\hsp{-0.3} system's\hsp{-0.3} pure-state\hsp{-0.3} step\hsp{-0.3} function.

To see why a quantum mixed state cannot be a step function of pure states, suppose we have $2$-qubit pure state \smash{$|\psi \rangle  \equiv a_1 |1\rangle \! +\! a_2 |2\rangle  \!+\! a_3 |3\rangle \! +\! a_4 |4\rangle$}, where \smash{$|1\rangle  \equiv |1,1\rangle $}, \smash{$|2\rangle  \equiv |1,2\rangle $}, etc., and \smash{$\sum\nolimits_{k = 1}^{\shiftmath{-0.7pt}{4}} {|a_k |^2  = 1}$}, where
\begin{equation}
a_k  \equiv \langle k|\psi \rangle
\label{eq:H.1}
\end{equation}
are \textit{wave-function overlaps} between pure quantum states $|\psi \rangle$ and $|k\rangle$.  The density matrix of $|\psi \rangle$ then has elements $\rho _{y,z}  \equiv \langle y|\psi \rangle \langle \psi |z\rangle  = a_y a_z^*$, so its mode-$1$ reduction is
\begin{equation}
\redx{\rho}{1} \scalebox{0.95}{$=\!\left(\!\! {\begin{array}{*{20}c}
   {\rho _{1,1}  \!+\! \rho _{2,2} } &\! {\rho _{1,3}  \!+\! \rho _{2,4} }  \\
   {\rho _{3,1}  \!+\! \rho _{4,2} } &\! {\rho _{3,3}  \!+\! \rho _{4,4} }  \\
\end{array}}\!\! \right)\!{\kern -2pt}=\!{\kern -2pt} \left(\!\! {\begin{array}{*{20}c}
   {a_1 a_1^*  \!+\! a_2 a_2^* } &\! {a_1 a_3^*  \!+\! a_2 a_4^* }  \\
   {a_3 a_1^*  \!+\! a_4 a_2^* } &\! {a_3 a_3^*  \!+\! a_4 a_4^* }  \\
\end{array}}\!\! \right)\!,$}
\label{eq:H.2}
\end{equation}
which is \textit{entirely} a function of the pure quantum wave function overlaps in \Eq{H.1}, so it is a true \textit{quantum mixture}.

In contrast, if we tried to create \Eq{H.2} from a \textit{time-average of a pure-state step function}, the decomposition states could still contain wave-function overlaps, but \textit{the mixture probabilities would be estimators of the classical probability that the system was actually in each particular pure quantum state}.

Therefore, we cannot truly prepare a system in the state of \Eq{2} as a time-averaged mixture, because the system would just be in different separable pure states at different times; a time-dependent \textit{pure} state.  \textit{In principle} (whether practical or not), one could guess how to assign measurements into subsets of the tomographic estimators to the pure state of the system at the exact time of measurement, and therefore determine the exact step function of the time-dependent pure state of preparation.

The \textit{quantum} mixture of \Eq{2} is different because its mixture probabilities inherit the \textit{instantaneous} nature of some pure parent state's superposition, since each is entirely{\kern -1pt} a{\kern -1pt} function{\kern -1pt} of{\kern -1pt} pure{\kern -1pt} wave-function{\kern -1pt} overlaps,{\kern -1pt} as{\kern -1pt} in{\kern -1pt} \Eq{H.2}.

Of course, we could simply \textit{purify} \Eq{2} and create that purified state in some larger system, and by focusing on the correct subsystem, we would then have prepared \Eq{2} as a true quantum mixture, but it would then \textit{not} be a \textit{time-averaged} mixture, and we could not claim it to have any true separability at any one time.

Note that even \textit{diagonal} quantum mixtures depend entirely on wave-function overlaps. For example, if $|\psi\rangle=$ \smash{$\frac{1}{\sqrt{2}}(|1\rangle+|4\rangle)$}, which is a Bell state so that \smash{$a_1 =a_4 =\frac{1}{\sqrt{2}}$} and \smash{$a_2 =a_3 =0$}, then \Eq{H.2} would become 
\begin{equation}
\redx{\rho}{1} =\left( {\begin{array}{*{20}c}
   {a_1 a_1^* } & 0  \\
   0 & {a_4 a_4^* }  \\
\end{array}} \right) = \left( {\begin{array}{*{20}c}
   {{\textstyle{1 \over 2}}} & 0  \\
   0 & {{\textstyle{1 \over 2}}}  \\
\end{array}} \right)\!,
\label{eq:H.3}
\end{equation}
which still depends only on the pure-state probability amplitudes of its pure parent state $|\psi\rangle$.

Thus, even though states like \Eq{2} have optimal decompositions into convex sums of pure product states, \textit{that is not indicative of the absence of entanglement}, because those decomposition states are not $k$-separable over the same $k$-partitions, and furthermore, the full state cannot be treated as instantaneously separable (meaning that it is never momentarily equal to one of its biseparable decomposition states).
\end{appendix}
\hypertarget{HedE}{}
%
\end{document}